\documentclass[journal]{IEEEtran}
\usepackage[nobreak]{cite}

\ifCLASSINFOpdf
\else
\fi
\usepackage{amsmath}
\ifCLASSOPTIONcompsoc
 \usepackage[caption=false,font=normalsize,labelfont=sf,textfont=sf]{subfig}
\else
 \usepackage[caption=false,font=footnotesize]{subfig}
\fi
\usepackage[dvipdfmx]{graphicx}
\usepackage{amsfonts}
\usepackage{amsthm}
\usepackage{bm}
\usepackage{bbm}
\usepackage[nolist, nohyperlinks]{acronym}
\usepackage{siunitx}
\usepackage{algorithm}
\usepackage{algpseudocode}
\usepackage{subfiles}
\usepackage[unicode,hidelinks]{hyperref}
\usepackage[active]{revise}
\usepackage{comment}
\usepackage{tikz}
\usetikzlibrary{calc,math,arrows,decorations.markings}
\usepackage{booktabs}
\usepackage{multirow}

\acrodef{ppm}{parts per milion; $10^{-6}$}
\acrodef{ASR}{automatic speech recognition}

\acrodef{DFT}{discrete Fourier transform}
\acrodef{STFT}{short-time Fourier transform}
\acrodef{GMM}{Gaussian mixture model}
\acrodef{CGMM}{complex Gaussian mixture model}
\acrodef{CC}{cross correlation}
\acrodef{CCF}{CC function}
\acrodef{GCC}{generalized cross correlation}
\acrodef{GCCF}{GCC function}
\acrodef{MCC}{multidimensional CC}
\acrodef{PHAT}{phase transform}
\acrodef{SCOT}{smoothed coherence transform}
\acrodef{ZP}{zero padding}
\acrodef{CRLB}{Cram\'er--Rao lower bound}
\acrodef{RMSE}{root mean square error}
\acrodef{MMSE}{minimum mean square error}
\acrodef{LWMA}{linear weighted moving average}
\acrodef{ML}{maximum likelihood}

\acrodef{W-DO}{W-disjoint orthogonality}
\acrodef{DOA}{direction of arrival}
\acrodef{DOAs}{directions of arrival}
\acrodef{RIR}{room impulse response}
\acrodef{ATF}{acoustic transfer function}
\acrodef{RTF}{relative transfer function}
\acrodef{TF}{time-frequency}
\acrodef{RPR}{relative phase ratio}
\acrodef{SCM}{spatial covariance matrix}
\acrodef{FSCM}{full-rank spatial covariance model}

\acrodef{BSS}{blind source separation}
\acrodef{ICA}{independent component analysis}
\acrodef{IVA}{independent vector analysis}
\acrodef{ILRMA}{independent low-rank matrix analysis}
\acrodef{VAD}{voice activity detection}
\acrodef{SS}{spectral subtraction}
\acrodef{DUET}{degenerate unmixing estimation technique}
\acrodef{SAE}{source activity estimation}
\acrodef{DNN}{deep neural network}
\acrodef{CNN}{convolutional neural network}
\acrodef{DC}{deep clustering}
\acrodef{NMF}{nonnegative matrix factorization}
\acrodef{MNMF}{multichannel nonnegative matrix factorization}
\acrodef{GSS}{golden-section search}
\acrodef{EM}{expectation--maximization}
\acrodef{MM}{majorization--minimization}
\acrodef{IBM}{ideal binary mask}

\acrodef{DS}{delay and sum}
\acrodef{MVDR}{minimum variance distortionless response}
\acrodef{MPDR}{minimum power distortionless response}
\acrodef{LCMV}{linearly constrained minimum variance}
\acrodef{MaxSNR}{maximum signal-to-noise ratio}
\acrodef{MWF}{multichannel Wiener filter}

\acrodef{DXC}{double-cross-correlation}
\acrodef{DXCP}{double-cross-correlation processor}
\acrodef{SSA}{spatial subtraction array}
\acrodef{SSA}{blind spatial subtraction array}

\acrodef{VM}{virtual microphone}
\acrodef{CGM}{complex-valued geometric mean}
\acrodef{MIN}{minimum value selection}
\acrodef{TFS}{time-frequency-bin-wise switching}
\acrodef{TFLC}{time-frequency-bin-wise linear combination}
\acrodef{AuxTDE}{auxiliary-function-based TDE}

\acrodef{SNR}{signal-to-noise ratio}
\acrodef{SDR}{signal-to-distortion ratio}
\acrodef{SIR}{signal-to-interference ratio}
\acrodef{SAR}{signal-to-artifacts ratio}
\acrodef{FAR}{false acceptance rate}
\acrodef{FRR}{false rejection rate}
\acrodef{SRDR}{signal-to-reconstruction distortion ratio}
\acrodef{MID}{mean absolute inconsistent delay}

\acrodef{ADMA}{asynchronous distributed microphone array}
\acrodef{DMA}{distributed microphone array}
\acrodef{SFM}{sampling frequency mismatch}
\acrodef{TDE}{TD estimation}
\acrodef{TD}{time delay}
\acrodef{STD}{sub-sample time delay}
\acrodef{TDOA}{time difference of arrival}
\acrodef{TOA}{time of arrival}

\acrodef{SiSEC}{community-based Signal Separation Evaluation Campaign}

\acrodef{JSPS}{Japan Society for the Promotion of Science}

\let\originalleft\left
\let\originalright\right
\renewcommand{\left}{\mathopen{}\mathclose\bgroup\originalleft}
\renewcommand{\right}{\aftergroup\egroup\originalright}

\newcommand{\naive}{na\"{i}ve }

\newcommand{\jw}{\mathbbm{i}\omega}

\newcommand{\ba}{\bm{a}}
\newcommand{\bx}{\bm{x}}

\newcommand{\tp}{^{\mathsf{T}}}
\newcommand{\htp}{^{\mathsf{H}}}
\newcommand{\inv}{^{\raisebox{.2ex}{$\scriptscriptstyle-\!$}1}}
\newcommand{\tpmat}{^{\!\!\!\mathsf{T}}}
\newcommand{\htpmat}{^{\!\!\!\mathsf{H}}}
\newcommand{\invmat}{^{\raisebox{.2ex}{$\!\!\!\scriptscriptstyle-\!$}1}}
\newcommand{\gt}{^{\star}}

\newcommand{\J}{\mathcal{J}}

\newcommand{\iu}{\mathbbm{i}}

\renewcommand{\proof}{{\noindent\bf Proof: }} 

\newtheorem{theorem}{Theorem}
\newtheorem{proposition}{Proposition}
\newtheorem{definition}{Definition}
\def\QED{~\rule[-1pt]{5pt}{5pt}\par\medskip}

\newcommand{\diag}{\mathop{\rm diag}}
\DeclareMathOperator*{\E}{\mathbb{E}}
\DeclareMathOperator*{\argmax}{arg~max}
\DeclareMathOperator*{\argmin}{arg~min}

\DeclareMathOperator{\sinc}{sinc}
\DeclareMathOperator{\rank}{rank}

\newcommand{\kHz}[1]{\SI{#1}{\kilo\hertz}}

\newcommand{\cm}[1]{\SI{#1}{\centi\meter}}
\newcommand{\m}[1]{\SI{#1}{\meter}}
\newcommand{\ms}[1]{\SI{#1}{\milli\second}}
\newcommand{\s}[1]{\SI{#1}{\second}}
\newcommand{\dB}[1]{\SI{#1}{\decibel}}
\newcommand{\pct}[1]{\SI{#1}{\percent}}

\newcommand{\secref}[1]{section~\ref{#1}}
\newcommand{\secsref}[1]{sections~\ref{#1}}
\newcommand{\Secref}[1]{Section~\ref{#1}}
\newcommand{\subsecref}[1]{subsection~\ref{#1}}

\newcommand{\tblref}[1]{Table~\ref{#1}}
\newcommand{\figref}[1]{Fig.~\ref{#1}} 

\newcommand{\figsref}[1]{Figs.~\ref{#1}} 
\newcommand{\subfigsref}[2]{\figsref{fig:#1}\subref{subfig:#2}}
\newcommand{\defref}[1]{Definition~\ref{#1}}
\renewcommand{\algref}[1]{Algorithm~\ref{#1}}

\newcommand{\T}{T}
\newcommand{\N}{N}
\newcommand{\bigs}{\mbox{\large $s$}}

\newcommand{\RTF}{\bg\uk\left(\ba\uk, \bt\right)}
\newcommand{\RTFh}{\bg\uk\htp\left(\ba\uk, \bt\right)}

\newcommand{\opt}{\ba\uk, \bt}
\newcommand{\RTFt}{\bg\uk\left(\bt\right)}
\newcommand{\RTFth}{\bg\uk\htp\left(\bt\right)}

\newcommand{\btheta}{\bm{\theta}}
\newcommand{\bV}{\bm{V}}
\newcommand{\bt}{\bm{\tau}}
\newcommand{\bbt}{\left(\bt\right)}
\newcommand{\er}{\bm{e}_r}
\newcommand{\bg}{\bm{g}}
\newcommand{\bsigma}{\bm{\sigma}}

\newcommand{\jwk}{\mathbbm{i}\omega_k}
\newcommand{\jwkt}{\mathbbm{i}\omega_k\tau}
\DeclareMathOperator{\Round}{round}
\newcommand{\vectorize}[2]{\left[{#1} \ \cdots \ {#2} \right]}

\newcommand{\kn}[1][]{_{{#1}kn}}
\newcommand{\uk}[1][]{_{{#1}k}}
\newcommand{\ik}{_{ik}}
\newcommand{\jk}{_{jk}}

\newcommand{\uij}{_{ij}}
\newcommand{\ijk}{_{ijk}}

\renewcommand{\algorithmicrequire}{\textbf{Input:}}
\renewcommand{\algorithmicensure}{\textbf{Output:}}
\algnewcommand{\Initialize}[1]{%
  \State \textbf{Initialize:}
  \State \hspace*{\algorithmicindent}\parbox[t]{0.8\linewidth}{\raggedright #1}
  \vspace{5pt}
}
\algnewcommand{\Lfunction}[3]{%
  \vspace{3pt}
  \State \textbf{Function} #1(#2)
  \State \hspace*{\algorithmicindent}\parbox[t]{0.8\linewidth}{\raggedright #3}
  \vspace{5pt}
  \State \textbf{end Function}
}
\newcommand{\mata}{
 \begin{pmatrix}
  a_{1k}e^{-\jwk\tau_{r1}} \\
  a_{2k}e^{-\jwk\tau_{r2}} \\
  \vdots \\[4pt]
  a_{Mk}e^{-\jwk\tau_{rM}}
 \end{pmatrix}
}

\newcommand{\loudspeaker}[3]{
 \coordinate (ls_A) at (#1, #2 + 0.3);
 \coordinate (ls_B) at (#1 + 0.3, #2 + 0.3);
 \coordinate (ls_C) at (#1, #2 + 0.8);
 \coordinate (ls_D) at (#1 + 0.3, #2 + 0.8);
 \coordinate (ls_E) at (#1 + 0.8, #2);
 \coordinate (ls_F) at (#1 + 0.8, #2 + 1.1);
 \filldraw[fill=gray] (ls_A) -- ($(ls_A)!1!#3:(ls_B)$) -- ($(ls_A)!1!#3:(ls_D)$) -- ($(ls_A)!1!#3:(ls_C)$) -- (ls_A) -- cycle;
 \filldraw[fill=gray] ($(ls_A)!1!#3:(ls_B)$) -- ($(ls_A)!1!#3:(ls_E)$) --($(ls_A)!1!#3:(ls_F)$) -- ($(ls_A)!1!#3:(ls_D)$) -- ($(ls_A)!1!#3:(ls_B)$) -- cycle;
}

\newcommand{\microphone}[2]{
 \coordinate (m_A) at (#1, #2);
 \coordinate (m_B) at (#1, #2 + 1);
 \filldraw[fill=gray] (#1 + 0.4, #2 + 0.5) circle[radius=0.4cm];
 \draw (m_A) -- (m_B);
}

\newcommand{\smallArrow}[3][]{
 \coordinate (ba_A) at #2;
 \coordinate (ba_B) at #3;
 \draw[#1,decoration={markings,mark=at position 1 with {\arrow[scale=1.3,>=stealth]{>}}},postaction={decorate}] (ba_A) -- (ba_B);
}
\newcommand{\smallBothArrow}[2]{
 \coordinate (bba_A) at #1;
 \coordinate (bba_B) at #2;
 \smallArrow{(bba_A)}{(bba_B)};
 \smallArrow{(bba_B)}{(bba_A)};
}

\begin{document}
%
\title{Estimation of Consistent Time Delays in Subsample via Auxiliary-Function-Based Iterative Updates}
%
%
%

\author{Kouei~Yamaoka,~\IEEEmembership{Student Member,~IEEE,}
    Yukoh~Wakabayashi,~\IEEEmembership{Member,~IEEE,}
	and~Nobutaka~Ono,~\IEEEmembership{Senior Member,~IEEE}
\thanks{K. Yamaoka, Y. Wakabayashi and N. Ono are with the Department of Computer Science, Graduate School of Systems Design, Tokyo Metropolitan University, 6-6 Asahigaoka, Hino, Tokyo 191-0065, Japan, e-mail: \{yamaoka-kouei@ed., wakayuko@, onono@\}tmu.ac.jp.}
}

\maketitle

\begin{abstract}
 In this paper, we propose a new algorithm for the estimation of multiple \acp{TD}.
 Since a \ac{TD} is a fundamental spatial cue for sensor array signal processing techniques, many methods for estimating it have been studied.
 Most of them, including generalized \ac{CC}-based methods, focus on how to estimate a \ac{TD} between two sensors.
 These methods can then be easily adapted for multiple \acp{TD} by applying them to every pair of a reference sensor and another one.
 However, these pairwise methods can use only the partial information obtained by the selected sensors, resulting in inconsistent \ac{TD} estimates and limited estimation accuracy.
 In contrast, we propose joint optimization of entire \ac{TD} parameters, where spatial information obtained from all sensors is taken into account.
 We also introduce a consistent constraint regarding \ac{TD} parameters to the observation model.
 We then consider a \ac{MCC} as the objective function, which is derived on the basis of \acl{ML} estimation.
 To maximize the \ac{MCC}, which is a nonconvex function, we derive the auxiliary function for the \ac{MCC} and design efficient update rules.
 We additionally estimate the amplitudes of the transfer functions for supporting the \ac{TD} estimation, where we maximize the Rayleigh quotient under the non-negative constraint.
 We experimentally analyze essential features of the proposed method and evaluate its effectiveness in \ac{TD} estimation.
 Code will be available at https://github.com/onolab-tmu/AuxTDE.
\end{abstract}

\acresetall

%
\IEEEpeerreviewmaketitle

\section{Introduction}
\label{sec:intro}
A \ac{TD} or \ac{TDOA} \cite{Jingdong06} observed between two sensors is a fundamental spatial cue for many signal processing techniques such as source localization, speech enhancement, and signal separation.
Source localization and \ac{DOA} estimation are essential techniques in audio \cite{Brandstein97,Gustafsson03,Xavier14,schwartz2014,LOCATA} and other various engineering fields including, sonar \cite{Carter:1981ke}, radar \cite{Protiva:2011io}, ground-penetrating radar \cite{lele14}, and reflection seismology \cite{Capon:1970it}. 
\ac{TD}-based localization methods are widely studied because of its importance \cite{Kien16,Kien17,yimao19}.
Speech enhancement is also important to extract an desired signal from noisy observation(s) \cite{makinoBSS2007,Gannot2017,Wakabayashi21}.
Resampling~\cite{Miyabe2015,Chinaev21} and synchronization~\cite{Coulson:2001kf} are also general topics, especially for asynchronous distributed systems \cite{Alexander11}.
For these techniques, the \acp{TD} are important spatial features, where any improvement in \ac{TDE} directly translates to their better performance, and numerous studies exist \cite{Jingdong06,Blandin12}.
\par
The mainstream \ac{TDE} techniques are based on the \ac{GCC} method \cite{knapp76}, which is the most popular technique.
Many techniques to improve the accuracy of \ac{GCC}-based \ac{TDE} have been proposed \cite{Chen:2004hs,Tashev:2009vj,Maximo20}, where many studies focus on how to estimate a single \ac{TD} between two different sensors.
This is because we can easily adapt these techniques for an array of more than three sensors, where there are multiple \acp{TD} to be estimated, by applying them repeatedly for every pair of a reference sensor and another one.
We call this simple solution the {\it pairwise method} in this paper.
\par
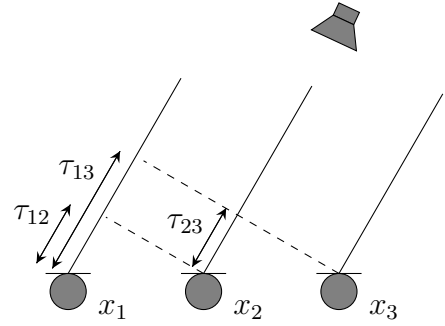
\begin{figure}
 \centering
  \begin{tikzpicture}[scale=0.6, rotate=-90]

   \microphone{8}{6}
   \microphone{8}{3}
   \microphone{8}{0}

   \coordinate (A) at (8, 0.5);
   \coordinate (B) at ($(A) + (5 * -0.866, 5 * 0.5)$);
   \coordinate (C) at (8, 3.5);
   \coordinate (D) at ($(C) + (4.8 * -0.866, 4.8 * 0.5)$);
   \coordinate (E) at (8, 6.5);
   \coordinate (F) at ($(E) + (4.6 * -0.866, 4.6 * 0.5)$);
   \draw (A) -- (B);
   \draw (C) -- (D);
   \draw (E) -- (F);

   \coordinate (CA) at ($(C) + (2.5 * -0.5, 2.5 * -0.866)$);
   \coordinate (EA) at ($(E) + (5 * -0.5, 5 * -0.866)$);
   \draw[dashed] (C) -- (CA);
   \draw[dashed] (E) -- (EA);

   \node at ($(A) + (0.8, 1)$) {\large $x_1$};
   \node at ($(C) + (0.8, 1)$) {\large $x_2$};
   \node at ($(E) + (0.8, 1)$) {\large $x_3$};

   \loudspeaker{2}{6.2}{-25}

   \coordinate (a) at ($(A) - (0.23, 0.7)$);
   \coordinate (b) at ($(a) + (1.5 * -0.866, 1.5 * 0.5)$);
   \coordinate (c) at ($(A) - (0.115, 0.35)$);
   \coordinate (d) at ($(c) + (3 * -0.866, 3 * 0.5)$);
   \coordinate (e) at ($(C) - (0.15, 0.25)$);
   \coordinate (f) at ($(e) + (1.5 * -0.866, 1.5 * 0.5)$);
   \smallBothArrow{(a)}{(b)}
   \smallBothArrow{(c)}{(d)}
   \smallBothArrow{(e)}{(f)}
   \node at (6.7, -0.3) {\large $\tau_{12}$};
   \node at (5.7, 0.7) {\large $\tau_{13}$};
   \node at (6.9, 3.1) {\large $\tau_{23}$};

  \end{tikzpicture}

 \caption{Time delays and a sensor array. The time delay $\tau_{13}$ is equal to the sum of time delays $\tau_{12}$ and $\tau_{23}$ physically. This relationship holds for any sensor array alignments.}
 \label{fig:array}
\end{figure}

Although the pairwise method is broadly used, we consider some problems.
First, the \ac{CC} between a sensor and another far from it may be low, which degrades the accuracy of a \ac{GCC}-based method.
This may be a serious problem, especially in a large-scale environment, e.g., distributed systems \cite{Alexander11}.
Second, \acp{TD} estimated by the pairwise method are {\it inconsistent}.
Assuming three sensors as shown in \figref{fig:array}, the \ac{TD} between sensors $1$ and $3$ is theoretically equal to the sum of those between sensors $1$ and $2$ and $2$ and $3$.
However, this does not hold true because of, for example, estimation errors caused by the presence of noise.
The best sensor to use as the reference sensor and how to identify it are unclear.
Summarizing the above, the pairwise method only uses partial information obtained from a pair of sensors.
For better \ac{TDE}, the spatial information obtained from all sensors should be taken into account.
In this paper, we thus aim to develop a new method for simultaneously estimating {\it consistent} \acp{TD} to improve their estimation accuracy.

\par
Let us consider observing a source signal with an $M$ channel sensor array.
The number of sensor pairs is $_MC_2$ ($M$ choose $2$), and the same number of \acp{TD} can be computed while only $M-1$ \acp{TD} exist theoretically; in other words, there is redundancy.
Some techniques utilize this redundancy to estimate a more accurate \ac{TD}, e.g., by introducing a multichannel \ac{CC} coefficient \cite{Jingdong03, Benesty04, Jingdong06}.
In contrast, we introduce a consistent constraint for estimating $M-1$ \acp{TD} and consider \ac{MCC} that encodes spatial cues obtained from all sensors.
Here, \ac{TDE} via the maximization of the \ac{MCC} has two difficulties: how to attain the \ac{TD} estimates with subsample precision and how to estimate them efficiently.
\par
The methodology of subsample \ac{TDE} is widely studied to improve the accuracy of \ac{GCC}-based estimation \cite{knapp76,Chen:2004hs,Tashev:2009vj}. 
A \naive \ac{TD} estimate is given by the location of the maximum of the discrete \ac{CC} function between two sensors.
Without further processing, the accuracy of a \ac{GCC}-based method is limited by the sampling frequency.
This can be a serious problem, especially for compact arrays.
For example, for a sound source radiated in an ambient atmosphere, the maximum \ac{TD} observed by two microphones spaced by \cm{4} is less than \ms{0.12}, i.e., less than two samples at \kHz{16}.
Such a sample level accuracy is insufficient for many applications, including, but not limited to, audio and optical\footnote{The terminology of ``time-of-flight'' is typically used instead.} processing techniques.
In the case of two sensors, interpolation is a popular and effective method to attain the subsample \ac{TD} estimate, where the \ac{CC} function is interpolated in the vicinity of the maximum.
The parabolic interpolation \cite{giovanni93} determines a quadratic function whose curve goes through three neighboring points, namely, the discrete maximum point of the \ac{GCC} function and its two adjacent points on both sides.
By using the vertex of the quadratic function instead of the discrete maximum, we can obtain a subsample \ac{TD} estimate.
Various schemes have been proposed, e.g., Gaussian curve fitting \cite{zhang06} and others \cite{francesco08,bo08,ran09}.
Yet another interpolation method is \acl{ZP} in the frequency domain, which corresponds to Dirichlet kernel interpolation \cite{Vetterli:2014vm}, where the ratio of nonpadded to padded signal lengths is the attainable subsample accuracy.
However, these techniques attain not the maximum of the \ac{GCC} funcion but an approximate one and are only applicable to the pairwise method.
\par
It is possible to find the maximum of the continuous \ac{GCC} function directly.
For band-limited signals, on the basis of the Nyquist--Shannon sampling theorem \cite{shannon49, ogawa06}, the continuous \ac{GCC} function is obtained by $\sinc$-interpolation of its discrete counterpart.
Its maximization is a nonconvex problem without a known closed-form solution.
Nevertheless, a locally optimal solution can be found with a search algorithm such as the exhaustive search scheme \cite{wang2016} and \ac{GSS} \cite{kiefer53}.
The \ac{GSS} is an efficient algorithm of ternary searches, and thus, it must be performed on a two-dimensional parameter space selected from $M-1$ \ac{TD} parameters in order.
The exhaustive algorithm can be applied to the search in $(M-1)$-dimensional parameter space; however, it is computationally demanding.
\par
In contrast, we previously proposed a technique of maximizing a continuous \ac{CC} function via the auxiliary-function-based iterative updates for subsample \ac{TDE} \cite{yamaoka2019WASPAA}.
This technique theoretically yields the same estimate as the exhaustive search, but the computational cost is markedly low owing to efficient updates.
By extending this method, in this paper, we propose an efficient algorithm, namely, \ac{AuxTDE}, for obtaining highly accurate and consistent \ac{TD} estimates.
First, we derive the objective function, i.e., the \ac{MCC} in the \ac{ML} sense.
Then, we show that the objective function can be globally bounded by a quadratic auxiliary function and can then be repeatedly maximized for guaranteed convergence to a local maximum.
Finally, we propose the \ac{AuxTDE} algorithm to estimate consistent \acp{TD}.
This method reaches the exact peak of the objective function, which means that the highly accurate subsample estimates can be obtained, and is the reference-free algorithm, which means that consistent \acp{TD} are obtained.
\par
The rest of this paper is organized as follows.
In \secref{sec:preliminary}, we define the signal model and briefly introduce the \ac{GCC}-method.
We also define what is consistent \acp{TD}.
In \secref{sec:formulation}, we formulate the estimation of consistent \acp{TD} and define the objective function.
In \secref{sec:aux_tdoa}, we first explain the fundamental idea and theories for our problem.
We here consider the case of two sensors as the simplest scenario, where we show the relationship between the objective function and \ac{CC}.
In \secref{sec:maux_tdoa}, we generalize the algorithm described in \secref{sec:aux_tdoa} and propose the technique for estimating consistent \acp{TD}.
Experimental analysis and numerical experiments are the topics in \secsref{sec:exp1} and \ref{sec:exp2}, respectively.
\Secref{sec:conclusion} concludes this paper.
\par
Note that this paper is partially based on conference paper \cite{yamaoka2019WASPAA} in which we proposed the technique introduced in \secref{sec:aux_tdoa}.
The contribution of this paper is that we extend the proposed method in \cite{yamaoka2019WASPAA} to multiple \acp{TD} and introduce the concept of consistent \acp{TD}. 


\section{Time delays estimation}
\label{sec:preliminary}
\subsection{Signal Model}
\label{subsec:model}
In this paper, we consider estimating every interchannel subsample \ac{TD} observed by an $M$ channel sensor array.
Let $x_{mkn}$ be the \ac{STFT} representation of the signal observed by the $m$th sensor at the discrete frequency $k$ in the $n$th time frame.
We model the observations as
\begin{align}
 \bx\kn &= \tilde{\bigs}\kn \bm{h}_k + \bm{u}\kn \label{eq:obs1} \nonumber \\
 &= \vectorize{x\kn[1]}{x\kn[M]}\tp, \\
 \bm{h}_k &= \vectorize{\tilde{a}_{1k}e^{-\jwk t_1}}{\tilde{a}_{Mk}e^{-\jwk t_M}} \nonumber \\
 &= \vectorize{h_{1k}}{h_{Mk}}\tp, \label{eq:ATF}
\end{align}
where $\tilde{\bigs}\kn$ is a source signal and $\bm{u}\kn = \vectorize{u\kn[1]}{u\kn[M]}\tp$ is noise signals at each sensor, where the superscript $\mathsf{T}$ denotes nonconjugate transposition.
$\mathbbm{i} = \sqrt{-1}$ denotes the imaginary unit, $\omega_k = 2\pi k / \T$ denotes the normalized angular frequency, and $\T$ denotes the number of samples in a frame.
$\bm{h}_k$ is the transfer function from the signal source to the sensor array, where $\tilde{a}_{mk}$ and $t_m$ denote the amplitude and the \ac{TOA}, respectively.
The \ac{TOA} represents the absolute time when the wave propagates to the sensor from the signal source.
We here consider the \ac{RTF} $\bm{g}_k$ \cite{doclo2015,shmulik2015}, which is defined as the ratio of the transfer function $\bm{h}_k$:
\begin{align}
 \RTF &= \bm{h}_k / h_{rk}  \nonumber \\
  &= \vectorize{a_{1k}e^{-\jwkt_1}}{a_{Mk}e^{-\jwkt_M}}\tp, \label{eq:RTF} \\
 \ba_k &= \vectorize{a_{1k}}{a_{Mk}}\tp, \\
 \bt &= \vectorize{\tau_{r1}}{\tau_{rM}}\tp,
\end{align}
where $a_{mk} = \tilde{a}_{mk}/\tilde{a}_{rk}$ is the frequency-dependent relative amplitude ($a_{mk} \in \mathbb{R}^+$), and $\tau_{rm} = t_m - t_r$ is the continuous \ac{TD} (\ac{TDOA}) between the reference sensor $r$ and the $m$th sensor.
Therefore, $a_{rk}=1$ for all $k$, and $\tau_{rr} = 0$.
This means that degree of freedom of the \acp{TD} is $M-1$ ($M$ \acp{TOA} minus one time origin).
Without loss of generality, we set to $r=1$ in this paper. 
Finally, the signal model \eqref{eq:obs1} with the \ac{RTF} is
\begin{align}
 \bx\kn &= \bigs\kn \RTF + \bm{u}\kn, \label{eq:obs}
\end{align}
where $\bigs\kn = \tilde{\bigs}\kn h_{rk}$ is the source image observed at the reference sensor.
Then, the objective of this paper is to estimate \acp{TD} $\bt$ from the observations $\bx\kn$.
In the rest of this paper, we denote scalars by regular letters and denote vectors and matrices by bold lower and upper case letters, respectively.

\subsection{GCC-based Time Delay Estimation}
\label{subsec:gcc}
The \ac{GCC} method \cite{knapp76,Tashev:2009vj} is commonly used for estimating a discrete \ac{TD} that maximizes the weighted \ac{CC} function
\begin{align}
 \Phi^{(rm)}(\tau) &= \frac{1}{\T}\sum_{k=-\T/2+1}^{\T/2} W_k S_k^{(rm)} e^{\jw_k \tau}, \label{eq:xspec} \\
 \hat{\tau}_{rm} &= \argmax_{\tau} \ \Phi^{(rm)}(\tau), \label{eq:gcc}
\end{align}
where $W_k \in \mathbb{R}^+$ is an arbitrary weight function for the \ac{GCC} function and $S_k^{(rm)}$ is the cross spectrum of $x_{mkn}$ and $x_{rkn}$.
Suitable weight functions have been proposed, e.g., \ac{GCC}-\acs{PHAT} (\acl{PHAT}) and \ac{GCC}-\acs{SCOT} (\acl{SCOT}):
\begin{align}
 W_k^{\rm PHAT} = |S_k^{(rm)}|^{-1}, \quad
 W_k^{\rm SCOT} = \left(S_k^{(rr)}S_k^{(mm)}\right)^{-\frac{1}{2}}.
\end{align}
The \ac{GCC} function with $W_k = 1$ is equivalent to the ordinary \ac{CC} function.
In typical implementations, the above \ac{GCC} function is only computed at discrete \acp{TD} given by the sampling frequency $F_s$ of the input signals, i.e., $\tau_{rm} \in \left\{ \frac{k}{F_s}\,|\,k=-\frac{\T}{2}+1,\ldots,\frac{\T}{2}\right\}$.
To improve the accuracy, it is necessary to remove this restriction, and many methods have been proposed as described in \secref{sec:intro}.
\par
The \ac{GCC} method is easily applicable to estimating multiple \acp{TD}; namely, we repeatedly obtain the \ac{GCC} function by computing \eqref{eq:xspec} and maximize it by solving \eqref{eq:gcc} for all $m$ except for $r$.
We call this approach, which uses only partial information, the pairwise method in this paper.
The algorithm of the pairwise method is quite simple, and many existing methods can be adapted; however, these \ac{TD} estimates computed by the pairwise method are basically inconsistent.

\subsection{Consistent Time Delays}
\label{subsec:consistency}
With $M$ sensors and one signal source, there are $M-1$ \acp{TD}.
This means that although we need to select one reference sensor to determine the absolute time origin, the theoretical \acp{TD} are independent of its selection.
In other words, the \acp{TD} should be consistent.
Now, we define what are consistent \acp{TD}. 

\begin{definition}[Consistent time delays]
 \label{def:consistency}
 Let $\tau_{rm}$ be the \acl{TD} between $m$th ($m=1, \dots, M$) and $r$th sensors. The \aclp{TD} are said to be consistent if
 \begin{align}
  \tau_{r'm} = \tau_{r'r} + \tau_{rm} \ \ \forall m, r, r', \label{eq:consistency}
 \end{align}
 where $r'$ is another reference sensor and $r' = 1, \dots, M$.
\end{definition}


\defref{def:consistency} is always satisfied when $\tau_{rm} = t_m - t_r$ in theory.
However, there is no guarantee that \acp{TD} estimated by the pairwise method satisfy \eqref{eq:consistency} since they estimate each \ac{TD} separately; hence, the \ac{TD} estimates are inconsistent.
In this paper, we model the observed signal \eqref{eq:obs} that based on the consistent \acp{TD}, i.e., $\tau_{rm} = t_m - t_r$.
Thus, by estimating all \acp{TD} simultaneously, those estimates naturally satisfy \eqref{eq:consistency}.

\section{Problem formulation}
\label{sec:formulation}
Starting from the signal model \eqref{eq:obs}, we aim to estimate consistent \acp{TD} in the \ac{ML} sense.
We here assume that the noise signals $\bm{u}\kn$ follow the complex multivariate Gaussian distribution $\mathcal{N}_c\left(\bm{\mu}, \bm{I}\right)$ with the mean $\bm{\mu} = \bm{0}$ and variance of an $M \times M$ identity matrix $\bm{I}$.
First, we eliminate the variable $\bigs\kn$ by replacing it with the \ac{ML} estimate $\hat{\bigs}\kn$.
Given $\RTF$, we obtain an \ac{ML} estimate of $\bigs\kn$ as follows:
\begin{align}
 \hat{\bigs}\kn &= \argmin_{s}\|\bx\kn - \bigs\RTF\|^2_2 \nonumber \\
 &= \frac{\bg\uk\htp\left(\ba\uk, \bt\right) \bx\kn}{\bg\uk\htp\left(\ba\uk, \bt\right) \bg\uk\left(\ba\uk, \bt\right)}, \label{eq:s}
\end{align}
where $\|\cdot\|_2^{}$ is the Euclidean norm of a vector and the superscript $\mathsf{H}$ denotes conjugate transposition.
We can also consider that $\bm{u}\kn$ follows the distribution $\mathcal{N}_c\left(\bm{\mu}, \bm{\Sigma}\right)$ with the channel-dependent variance $\bm{\Sigma}$; $\bm{\Sigma}$ is an $M \times M$ diagonal matrix whose $m$th diagonal entry is $\sigma_m^2$.
In this case, by using the following weighted vectors instead of the original ones, we obtain the same \ac{ML} estimate $\hat{\bigs}\kn$:
\begin{align}
  \bx'\kn &=  \diag \left(\bsigma\right)\inv \bx\kn, \label{eq:xs} \\
  \ba'\uk &=  \diag \left(\bsigma\right)\inv \ba\uk, \label{eq:gs}
\end{align}
where $\diag(\cdot)$ is a function that returns a square diagonal matrix with the elements of an input vector and $\bsigma = \vectorize{\sigma_1}{\sigma_M}$.
Hereafter, we basically consider the channel-dependent case and omit prime marks of $\bx'\kn$ and $\ba'\uk$ for notational ease.
\par
The optimal $\hat{\bigs}\kn$ is the function of the amplitudes $\ba\uk$ and \acp{TD} $\bt$.
Now we substitute $\hat{\bigs}\kn$ to \eqref{eq:obs} and find \ac{ML} estimates of $\ba\uk$ and $\bt$.
\begin{gather}
  \argmin_{\opt} \ \frac{1}{\T}\sum_{k=-\T/2+1}^{\T/2}\J\uk^{(1)}\left(\opt\right) \nonumber\\
 \mathrm{s.t.} \ a_{mk} \geq 0, a_{rk} = 1, \tau_{rr} = 0 \ \forall m, k, \\
   \J\uk^{(1)}\left(\opt\right) = \E\left[\Bigl\|\bx\kn - \frac{\bg\htp\uk\left(\ba, \bt\right)\bx\kn}{\bg\htp\uk\left(\ba, \bt\right)\RTF}\RTF\Bigr\|^2_2\right], \label{eq:min_J1}
\end{gather}
where $\E[\cdot]$ denotes the expectation operator, which is replaced by the time average in practice by assuming the ergodic process.
By expanding \eqref{eq:min_J1} and ignoring constant terms that include neither $\ba\uk$ nor $\bt$, the minimization of the sum of $\J\uk^{(1)}\left(\opt\right)$ is reduced to the maximization of the following objective function $\J\left(\opt\right)$:
\begin{align}
 \argmax_{\opt} \ &\J\left(\opt\right) \ \ 
 \mathrm{s.t.} \ a_{mk} \geq 0, \tau_{rr} = 0 \ \forall m, k, \label{eq:opt_maux}
\end{align}\vspace{-15pt}%
\begin{align}
 \J\left(\opt\right) &= \frac{1}{\T}\sum_{k=-\T/2+1}^{\T/2}\J\uk\left(\opt\right), \label{eq:obj_maux} \\
 \J\uk\left(\opt\right) &= \ \frac{\RTFh \bV\uk \, \RTF}{\ba\uk\tp \ba\uk}, \label{eq:obj_k_maux} \\
 \bV\uk &= \ \E\left[\bx\kn \bx\htp\kn \right], \label{eq:scm}
\end{align}
where we use the relationship of $\RTFh\RTF = \ba\uk\tp\ba\uk$, computed form \eqref{eq:RTF}, for the denominator.
Since the objective function \eqref{eq:obj_maux} consists of the \ac{CC} functions for every pair of sensors (which comes from the covariance matrix \eqref{eq:scm}), we call this objective function \ac{MCC}.
Additionally, the \ac{MCC} is invariant to the scale of the amplitudes $\ba\uk$, i.e., $\J\uk(\ba\uk, \bt)  = \J\uk(\gamma\uk\ba\uk, \bt)$ with any real numbers $\gamma\uk$.
Hence, the constraint $a_{rk} = 1$ (actually, $a'_{rk} = \sigma_r$ due to \eqref{eq:gs}) can be satisfied by the normalization at the end of the optimization sequences.
Finally, our goal is to estimate the subsample \acp{TD} $\bt$ that maximize the \ac{MCC} \eqref{eq:obj_maux}.
We also estimate amplitudes $\ba\uk$, which may contribute to improving the accuracy of \ac{TDE}.
\par
In \secref{sec:aux_tdoa}, we first estimate $\bt$ in the case of $M=2$ as the simplest scenario (i.e., we estimate only one \ac{TD}), where we introduce the theories essential for solving this optimization problem and obtaining the highly accurate subsample \ac{TD} estimate.
In \secref{sec:maux_tdoa}, we then generalize the algorithm described in \secref{sec:aux_tdoa} in the case of $M$ sensors.


\section{subsample time delay estimation}
\label{sec:aux_tdoa}
In this section, we explain the technique of maximizing a \ac{GCC} function by the auxiliary function method as the special case of the proposed method, which was originally presented in the conference paper \cite{yamaoka2019WASPAA}.

\subsection{Problem Formulation}
\label{subsec:problem_aux}
In this section, we consider estimating a subsample \ac{TD} between two observed signals.
For simplicity, we here assume that the amplitude $a_{mk}$ is $1$ for all elements and the variances $\sigma_m^2$ are common for all sensors.
In this case, the \ac{RTF} \eqref{eq:RTF} is
\begin{align}
 \bg\uk\left(\tau_2\right) &= \left[1, e^{-\jwk \tau_{21}} \right]\tp,
\end{align}
where the reference sensor is set to $r=1$.
For notational ease, we hereafter use $\tau=\tau_{21}$ in this section.
Then, the objective function \eqref{eq:obj_k_maux} at the $k$th frequency becomes
\begin{align}
 \J\uk\left(\tau\right) &= V_{12k} e^{-\jwkt} + V_{21k} e^{\jwkt},
\end{align}
where $V\ijk$ is the $(i, j)$ element of the Hermitian matrix $\bV\uk$, and we omit the constant terms that do not include $\tau$.
Then, by using the conjugate relationship of the first and second terms, we obtain the following optimization problem \eqref{eq:opt_maux}:
\begin{align}
  \argmax_{\tau} &\ \J\left(\tau\right), \label{eq:opt_aux} \\
  \J\left(\tau\right) =& \frac{2}{\T}\sum_{k=-\T/2+1}^{\T/2} V_{21k} e^{\jw_k \tau}. \label{eq:Phi}
\end{align}
Equation \eqref{eq:Phi} exactly means the inverse \ac{DFT} of the cross spectrum, i.e., the \ac{CC} function between the real discrete signals observed by the sensors, $x\kn[1]$ and $x\kn[2]$, in the time domain.
From the above, the optimization problem \eqref{eq:opt_maux} reduces to the maximization of the \ac{CC} function.
Note that the relevance between the maximization of the \ac{CC} function and estimation of the \ac{TD} on the basis of \ac{ML} estimation was also discussed in \cite{wang2016}.
\par
Now, we consider finding a continuous variable $\tau \in \mathbb{R}$ maximizing the continuous function \eqref{eq:Phi}.
We can rewrite \eqref{eq:Phi} as a sum of cosine functions using the conjugate symmetry of $V_{21k}$ and Euler's formula,
\begin{align}
  \J\left(\tau\right) = \frac{1}{\T} \sum_{k=0}^{\T/2} A_k\cos (\omega_k \tau + \phi_k), \label{eq:obj_aux}
\end{align}
where $A_k = 2\beta_k |V_{21k}|$, $\phi_k = \angle V_{21k}$, $\angle$ takes the phase of input argument, and $\beta_0 = \beta_{\T/2} = 1$ and $\beta_k=2$ for $k\not\in \{0,\T/2\}$.
The arbitrary weight function $W_k$ can be introduced by replacing the definition of $A_k$ with $A_k = 2\beta_k |W_k V_{21k}|$.
In this case, the objective function corresponds to the \ac{GCC} function.
Our goal is thus to compute the \ac{TD} estimate with subsample precision by maximizing \eqref{eq:obj_aux}.

\subsection{Auxiliary Function Method for Subsample TDE}
\label{subsec:adapt_aux}
The auxiliary function method (also known as the \ac{MM} algorithm \cite{David04,Kenneth2016}) is well known as the generalization of the \textit{expectation--maximization} algorithm and applied to various algorithms \cite{Lee:2001ti,Kameoka:2007fa,ono2011,Andreas21}.
For adapting it to our problem, we would like to find an auxiliary function $Q_s(\tau, \btheta)$ such that
\begin{itemize}
    \item $\J(\tau) \geq Q_s(\tau, \btheta)$ for any $\tau$ and $\btheta$,
    \item for any $\tau$, $\exists \btheta = f(\tau)$ such that $\J(\tau) = Q_s(\tau,\btheta)$,
\end{itemize}
where $\btheta=(\theta_0, \theta_1, \cdots, \theta_{\T/2})$ are auxiliary variables.
Provided such a $Q_k(\tau,\btheta)$ exists and given an initial estimate $\hat{\tau}^{(0)}$, the following sequence of updates is guaranteed to converge to a local maximum:
\begin{align}
    \btheta^{(\ell)} = f(\hat{\tau}^{(\ell)}), \quad
    \hat{\tau}^{(\ell+1)} = \underset{\tau \in \mathbb{R}}{\argmax} \ Q_s(\tau, \btheta^{(\ell)}),
\end{align}
where $\ell$ is the iteration index.

\subsection{Quadratic Auxiliary Function for Continuous GCC}
\label{sec:QAFS}
\begin{figure}[t]
 \centering
 \includegraphics[width=1\columnwidth]{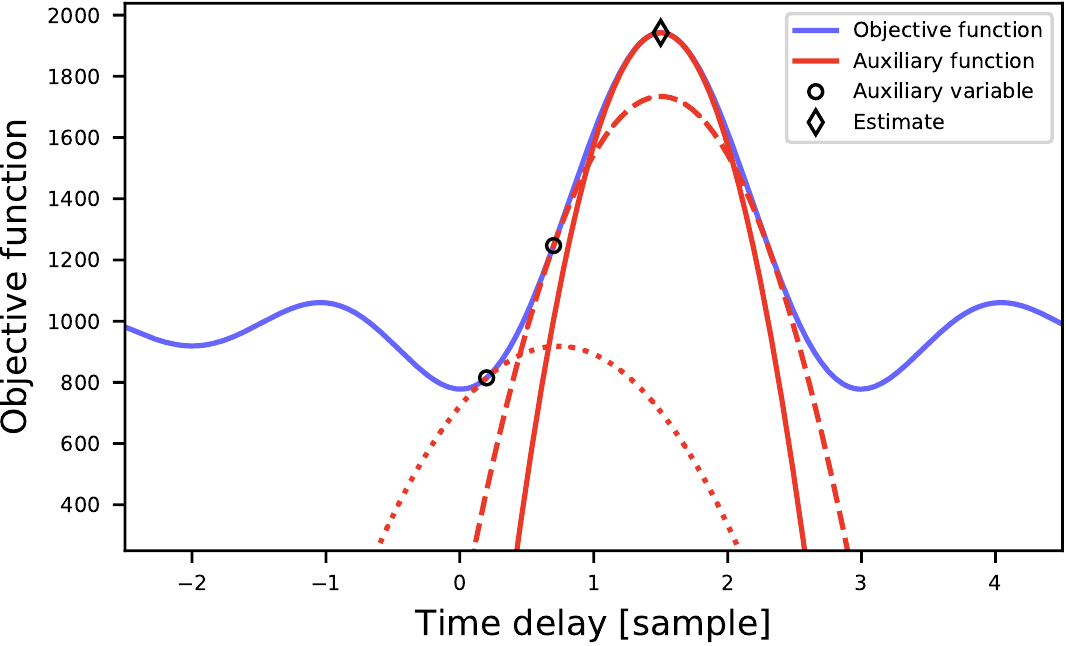}
  \caption{Examples of the objective function \eqref{eq:obj_aux} and the auxiliary function \eqref{eq:aux}. The objective function \eqref{eq:obj_aux} is the \ac{CC} function, which is a strictly unimodal function around the global maximum. Given an initial estimate picked from such a unimodal period, the proposed method must converge to the global optimal solution with monotonic increases in the objective function.}
 \label{fig:aux}
\end{figure}

This section provides a quadratic auxiliary function for $\J(\tau)$, i.e., $\J(\tau) \geq Q_s(\tau, \btheta)$.
\begin{theorem}
\label{th:aux}
The following is an auxiliary function for $\J(\tau)$,
\begin{align}
 Q_s(\tau, \btheta) = \frac{1}{\T} \sum_{k=0}^{\T/2} \Biggl\{ &- \frac{A_k}{2} \frac{\sin \theta_k}{\theta_k}(\omega_k \tau + \phi_k + 2\nu_k\pi)^2 \nonumber \\
 &+ \cos \theta_k + \frac{1}{2}\theta_k \sin \theta_k \Biggr\} \label{eq:aux},
\end{align}
where $\nu_k \in \mathbb{Z}$ is such that $|\omega_k \tau + \phi_k + 2\nu_k \pi | \leq \pi$.
The auxiliary variables are $\theta_k$ and $\nu_k$, then $Q_s(t, \btheta) = \J(\tau)$ when
\begin{align}
 \theta_k = \omega_k \tau + \phi_k + 2\nu_k\pi. \label{eq:theta}
\end{align}
\end{theorem}
This theorem is a direct consequence of the following inequality for a cosine function, which is of general interest.
\begin{proposition}
\label{prop:cos}
Let $|\theta_0|\leq \pi$. For any real number $\theta$, the following inequality is satisfied:
\begin{align}
    \cos \theta \geq -\frac{1}{2}\frac{\sin \theta_0}{\theta_0}\theta^2 + \left(\cos \theta_0 + \frac{1}{2}\theta_0 \sin \theta_0\right). \label{eq:ineq}
\end{align}
When $|\theta_0|<\pi$, equality holds if and only if $|\theta| = |\theta_0|$. 
When $|\theta_0|=\pi$, equality holds if and only if $\theta=(2\nu+1)\pi$, $\nu\in \mathbb{Z}$. 
\end{proposition}
Proof of Theorem \ref{th:aux} and Proposition \ref{prop:cos} are presented in the appendix.
\figref{fig:cos_aux} shows examples of the auxiliary functions for a cosine function \eqref{eq:ineq} and \figref{fig:aux} is for the objective function \eqref{eq:obj_aux}.

\begin{figure}[t]
 \centering
 \includegraphics[width=1\columnwidth]{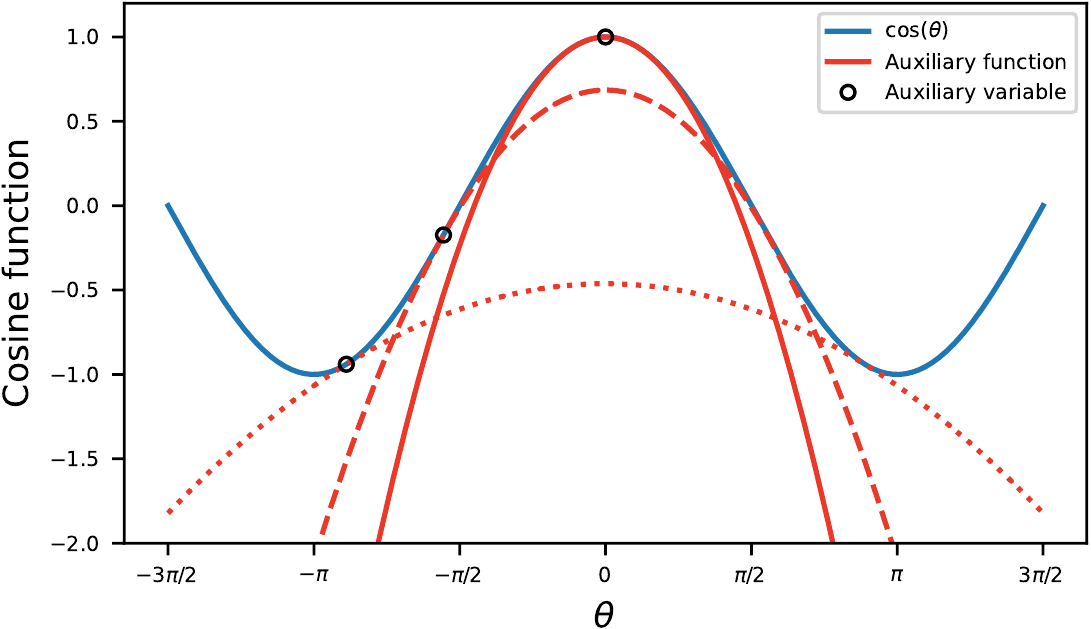}
 \caption{Example of the quadratic auxiliary function for the cosine function, where the point of tangency is the auxiliary variable $\theta_k$ at the $k$th frequency. Although other surrogate functions should exist, e.g., shown in \cite{Kenneth2016}, the proposed function is always the best for convergence, which requires only one update except for $\theta_k = \pm \pi$.}
 \label{fig:cos_aux}
\end{figure}

\subsection{Derivation of Auxiliary Function and Update Rules}
Since $Q_s(\tau, \btheta)$ is a quadratic function, it is easily maximized with respect to $\tau$ by setting its derivative to zero:
\begin{align}
\frac{\partial Q_s(\tau, \btheta)}{\partial \tau} = - \sum_{k=0}^{\T/2} A_k \omega_k \frac{\sin \theta_k}{\theta_k}(\omega_k \tau + \phi'_k) = 0,
\end{align}
where $\phi'_k = \phi_k + 2\nu_k \pi$,
Therefore, the maximizer is
\begin{align}
    \hat{\tau} &=\frac{\sum_{k=0}^{\T/2} A_k \omega_k^2 (\sin\theta_k/\theta_k) (- \phi'_k / \omega_k)}{\sum_{k=0}^{\T/2} A_k \omega_k^2 (\sin\theta_k/\theta_k)}. \label{eq:t}
\end{align}
Now, under the condition of equality~\eqref{eq:theta}, we can substitute $\phi'_k = \theta_k - \omega_k \tau$ for \eqref{eq:t} and obtain the final update rules:
%
\begin{align}
 \nu_k^{(\ell)} & \gets \Round\left\{ - \left(\omega_k \tau^{(\ell)} + \phi_k \right) / 2\pi \right\} , \label{eq:upd_n} \\
 \theta_k^{(\ell)} & \gets \omega_k \tau^{(\ell)} + \phi_k + 2 \nu_k^{(\ell)} \pi,\quad k=0,\ldots,\frac{\T}{2}, \label{eq:upd_theta} \\
 \tau^{(\ell+1)} & \gets \tau^{(\ell)} - \frac{\sum_{k=0}^{\T/2}A_k\omega_k^2\left(\sin\theta_k^{(\ell)}/\theta_k^{(\ell)}\right)\frac{\theta_k^{(\ell)}}{\omega_k}}{\sum_{k=0}^{\T/2}A_k\omega_k^2\left(\sin\theta_k^{(\ell)}/\theta_k^{(\ell)}\right)}, \label{eq:upd_tau}
\end{align}
where $\Round[\cdot]$ is a function that rounds the input to the nearest integer.
Interestingly, the second term of \eqref{eq:upd_tau} is a weighted sum of the auxiliary variables scaled by the normalized angular frequency, i.e., $\theta^{(\ell)}_k/\omega_k$, and $\theta^{(\ell)}_k/\omega_k$ corresponds to the candidate of the \ac{TD} estimate at the frequency $k$.



\section{AuxTDE: Consistent time delay estimation}
\label{sec:maux_tdoa}
In this section, we propose \ac{AuxTDE}, the method for estimating multiple \acp{STD} simultaneously.
This is the generalization of the method introduced in the previous section.

\subsection{Technical Approach for TDE}
\label{subsec:app}
As we mentioned in \secref{sec:formulation}, we aim to maximize the objective function \eqref{eq:obj_maux} to obtain subsample \ac{TD} estimates.
To solve the joint optimization problem \eqref{eq:opt_maux}, we first consider to optimize $\bt$ with fixed $\ba_k$.
Then, the denominator of \eqref{eq:obj_maux} is constant that does not include $\bt$.
Now, the question is how to maximize the sum of numerator with respect to $\bt$, which has the following structure:
\begin{align}
 \RTFth \bV\uk \, \RTFt = \mata\htpmat \bV\uk \mata,
\end{align}
where the parameters to be optimized are the exponent.
Unfortunately, the maximization of the above function has no closed-form solution.
Therefore, we consider applying the auxiliary function method as in \secref{sec:aux_tdoa}.
\par
The objective function at the $k$th frequency \eqref{eq:obj_k_maux} can be rewritten as a sum of the cosine function using the conjugate symmetry of $\bV\uk$ and Euler's formula, the same as the derivation of \eqref{eq:obj_aux},
\begin{align}
 \J\uk\left(\bt\right) &= \sum_{i=1}^{M}\sum_{j=1}^{M} A\ijk \cos \left( \omega\uk \tau_{ij} + \phi\ijk \right), \label{eq:obj_mcos}
\end{align}
where $A\ijk = \beta_k a\ik a\jk |V\ijk|$, $\phi\ijk = \angle V\ijk$, and the denominator is omitted.
It is worth noting that because the difference between two \acp{TD}, i.e., $\tau_{ij} = \tau_{rj} - \tau_{ri}$ in \eqref{eq:obj_mcos}, is equal to $t_j - t_i$, the objective function is independent of the reference sensor $r$.
Now, we propose the quadratic form auxiliary function for \ac{MCC}, which has the closed-form solution.
Here, we consider updating $\bt$, $\ba\uk$, and $\sigma_m$ alternately.

\subsection{Quadratic Form Auxiliary Function for TDE}
Similarly to \subsecref{subsec:adapt_aux}, we design the auxiliary function $Q(\bt, \btheta)$ for multiple \acp{TD} $\bt$ that satisfies the following properties:
\begin{itemize}
    \item $\J\bbt \geq Q(\bt, \btheta)$ for any $\bt$ and $\btheta$,
    \item For any $\bt$, $\exists \btheta = f(\bt)$ such that $\J(\bt) = Q(\bt,\btheta)$,
\end{itemize}
where $\btheta = (\theta\ijk)_{1 \leq i,j \leq M, 0 \leq k \leq \T/2}$ are auxiliary variables.
Provided such a $Q(\bt,\btheta)$ exists and given an initial estimate $\hat{\bt}^{(0)}$, the following sequence of updates is guaranteed to converge to a local maximum:
\begin{align}
    \btheta^{(\ell)} = f(\hat{\bt}^{(\ell)}), \quad
    \hat{\bt}^{(\ell+1)} = \underset{\bt\in\mathbb{R}}{\arg\max} \ Q(\hat{\bt}^{(\ell)}, \btheta^{(\ell)}).
\end{align}
\par
\begin{figure}[t]
 \begin{minipage}{0.49\hsize}
  \centering
  \includegraphics[width=1\columnwidth]{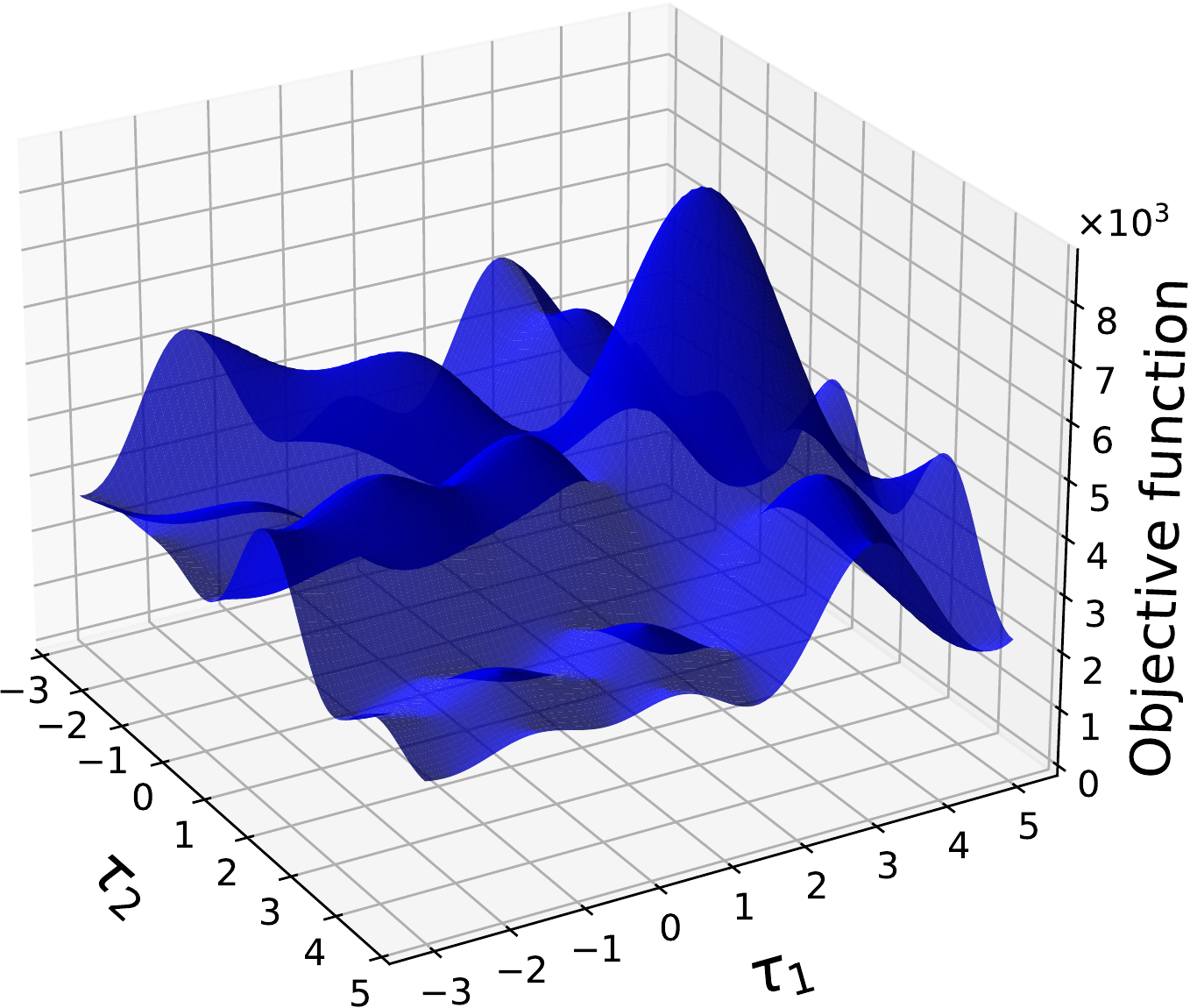}\\[-10pt]
  \subfloat[\label{subfig:mobj}]{\hspace{0.01\linewidth}}
 \end{minipage}
 \begin{minipage}{0.49\hsize}
  \centering
  \includegraphics[width=1\columnwidth]{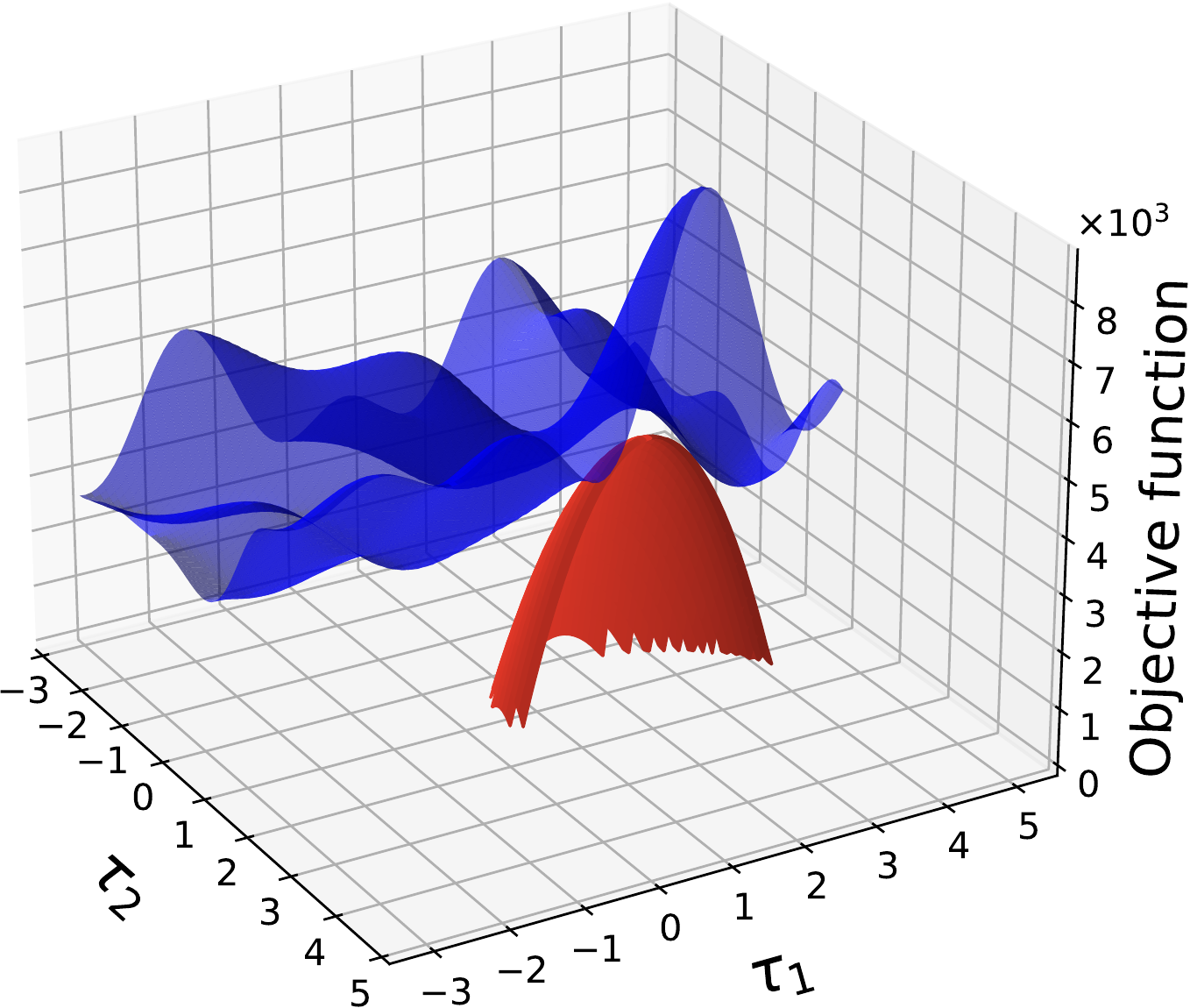}\\[-10pt]
  \subfloat[\label{subfig:maux1}]{\hspace{0.01\linewidth}}
 \end{minipage}
 \\
  \begin{minipage}{0.49\hsize}
  \centering
  \vspace{10pt}
  \includegraphics[width=1\columnwidth]{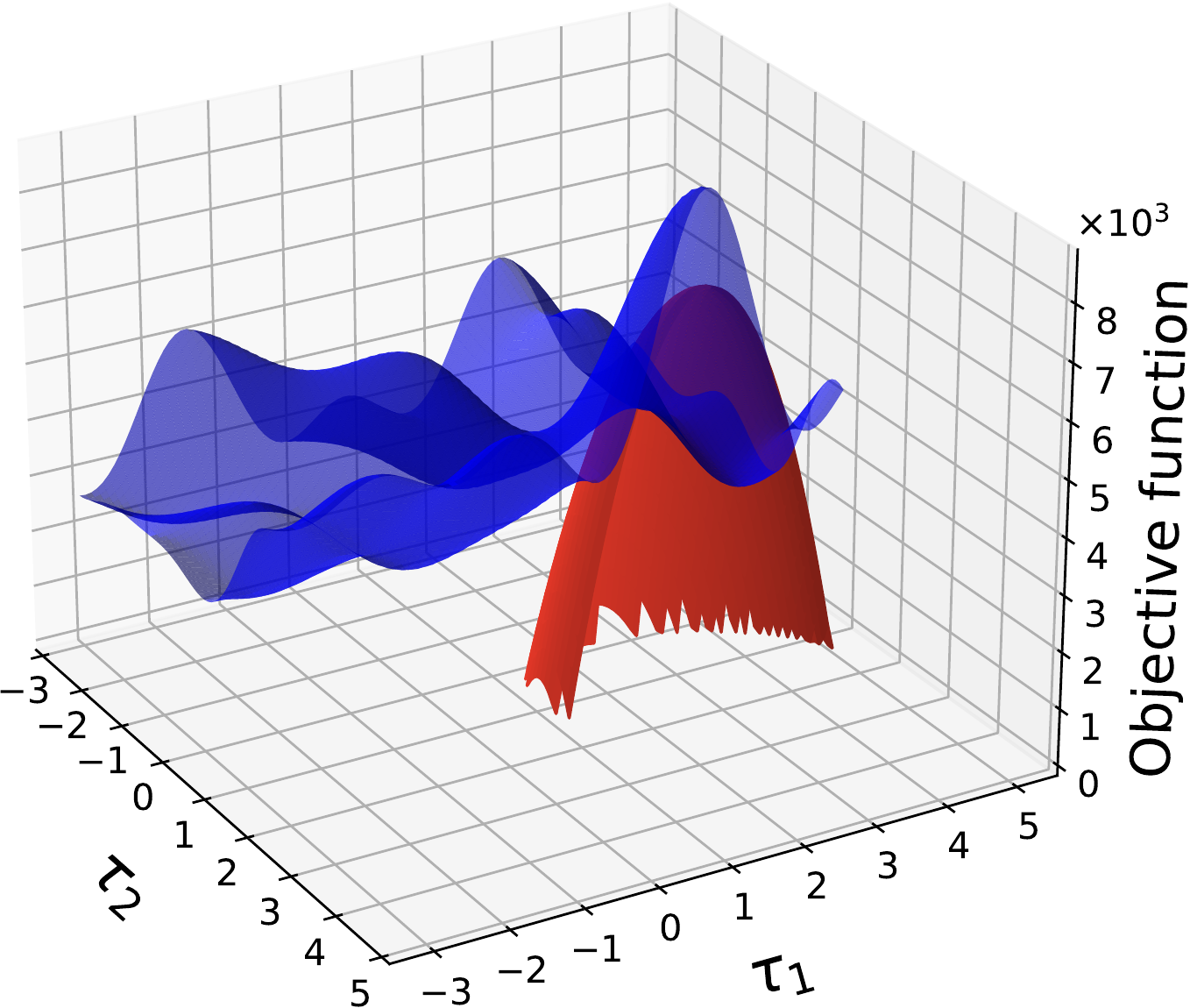}\\[-10pt]
  \subfloat[\label{subfig:maux2}]{\hspace{0.01\linewidth}}
 \end{minipage}
 \begin{minipage}{0.49\hsize}
  \centering
  \vspace{10pt}
  \includegraphics[width=1\columnwidth]{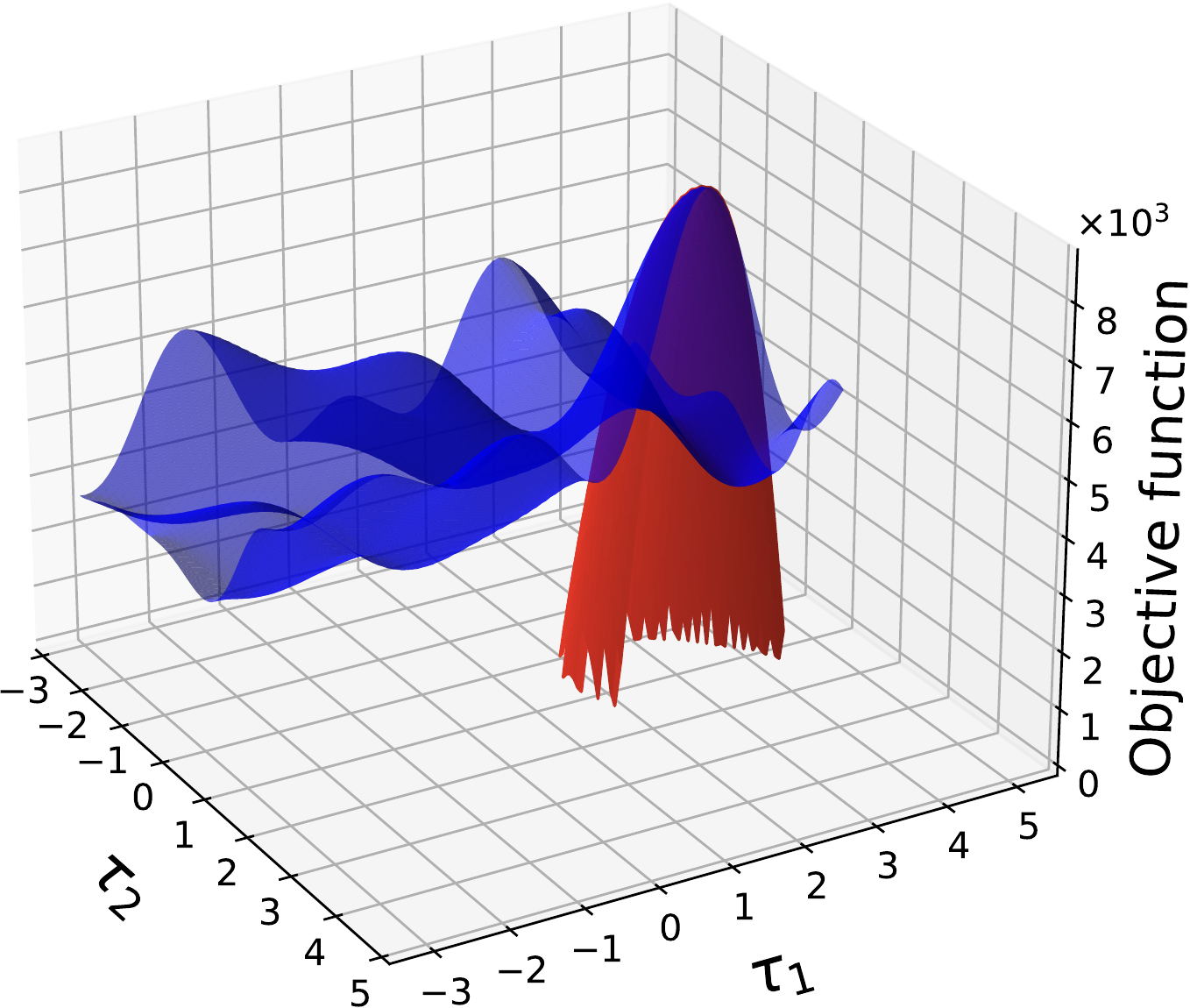}\\[-10pt]
  \subfloat[\label{subfig:maux3}]{\hspace{0.01\linewidth}}
 \end{minipage}
 \caption{Examples of the objective function (blue) and auxiliary functions (red), where these functions are shown in only the range that takes a value more than $0$. (a) objective function only. (b)--(d) auxiliary functions, where the \ac{TD} estimates are $(1.7, 1.5)$, $(2, 1.5)$, and $(3, 1.5)$, respectively.}
  \label{fig:maux}
\end{figure}

We here propose an auxiliary function for $\J(\bt)$, i.e., $\J(\bt) \geq Q(\bt, \btheta)$.
\begin{theorem}
\label{th:maux}
The following is an auxiliary function for $\J(\bt)$:
\begin{align}
 Q(\bt, \btheta) &= \frac{1}{\T}\sum_{k=0}^{\T/2} Q\uk(\bt, \btheta), \label{eq:Q} \\
 Q\uk(\bt,\btheta) &= \sum_{i=1}^{M}\sum_{j=1}^{M}
 \begin{aligned}[t]
  \Biggl\{ &- B\ijk \left(\omega\uk \tau\uij + \phi\ijk + 2\nu\ijk\pi\right)^2 \\
  &+ \cos \theta\ijk + \frac{1}{2}\theta\ijk \sin \theta\ijk \Biggr\},
 \end{aligned} \label{eq:Qk_sigma} \\
 B\ijk &= \frac{A\ijk}{2} \frac{\sin\theta\ijk}{\theta\ijk}, \label{eq:bijk}
\end{align}
where $\nu\ijk \in \mathbb{Z}$ is such that $|\omega\uk \tau_{ij} + \phi\ijk + 2\nu\ijk\pi| \leq \pi$.
 The auxiliary variables are $\theta\ijk$ and $\nu\ijk$ and $Q(\bt, \btheta) = \J\bbt$ holds when
\begin{align}
 \theta\ijk = \omega\uk \tau_{ij} + \phi\ijk + 2\nu\ijk\pi, \ \forall i,j,k. \label{eq:equality}
\end{align}

\end{theorem}
This theorem is a direct consequence of Proposition~\ref{prop:cos} with regards to a cosine function.
The auxiliary function \eqref{eq:Qk_sigma} can be rewritten as the vector quadratic form:
\begin{align}\hspace{-5pt}
 Q\uk(\bt, \btheta) &= - \bt\tp \bm{C}\uk \bt - 2\bm{c}\uk\tp \bt + {\rm Const.}, \label{eq:Qk} \\
 \bm{C}\uk &= \omega\uk^2 \left\{ \diag\left( \bm{b}\uk \right) - \bm{B}\uk \right\}, \label{eq:C} \\
 \bm{b}\uk &= \left(\sum_{i=1}^{M}B\uk[i1], \ \dots, \ \sum_{i=1}^{M}B\uk[iM] \right)\tpmat, \label{eq:b} \\
 \bm{c}\uk &= \omega\uk\left(\sum_{i=1}^{M}B\uk[i1]\phi'\uk[i1], \ \dots, \ \sum_{i=1}^{M}B\uk[iM]\phi'\uk[iM] \right)\tpmat, \label{eq:c}
\end{align}
where $\bm{B}\uk = (B_{ij})_{k, 1 \leq i,j \leq M}$ and  $\phi'\ijk = \phi\ijk + 2 \nu\ijk\pi$.
$\bm{C}\uk$ is the positive semidefinite matrix \cite{madison2015} and thus $-\bt\tp \bm{C}\uk \bt$ is the convex upward function.
We show an example of the objective function and auxiliary functions in \subfigsref{maux}{mobj}--\subref{subfig:maux3}, where $M=3$ (that is, there are two \ac{TD}s).
The objective function, the \ac{MCC}, is globally lower bounded by the proposed auxiliary function $Q(\bt, \btheta)$ at any point and has exactly one point of tangency.

\subsection{Derivation of Auxiliary Function and Update Rules}
To maximize the auxiliary function $Q(\bt, \btheta)$ under the constraint of $\tau_{rr} = 0$, we find the stationary point of the function $\sum_{k=0}^{\T/2} Q\uk(\bt, \btheta) - \lambda\tau_{rr}$, where $\lambda$ is the Lagrange multiplier.
Since $Q\uk(\bt, \btheta)$ is a quadratic form, it is easily maximized with respect to $\bt$ by setting its derivative to zero:
\begin{align}
  \frac{\partial Q\left(\bt, \btheta\right)}{\partial \bt\tp} &= - \sum_{k=0}^{\T/2} \left\{ 2 \bm{C}\uk \bt + 2\bm{c}\uk\right\} - \lambda\er = 0, \label{eq:partial}
 \end{align}
 where $\er$ is a unit vector whose only $r$th element is $1$.
Then, we obtain
  \begin{align}
  \bar{\bt} &= - \left( \sum_{k=0}^{\T/2} \bar{\bm{C}}\uk \right)\invmat \sum_{k=0}^{\T/2} \bar{\bm{c}}\uk, \label{eq:tau}
  \end{align}
where $\bar{\bm{C}}\uk$ is an $M-1 \times M-1$ matrix eliminating the $r$th row and column from $\bm{C}_k$, namely, $\bar{\bm{C}}\uk = (C_{ij})_{k, 1 \leq i,j \leq M, i,j \neq r}$.
Similarly, $\bar{\bt} = (\tau_{ri})_{1 \leq i \leq M, i \neq r}$ and $\bar{\bm{c}}\uk = (c_i)_{1 \leq i \leq M, i \neq r}$.
\par
Finally, under the condition of equality \eqref{eq:equality}, we can substitute $\phi'\ijk = \theta\ijk - \omega\uk \tau_{ij}$ and obtain the update rules:
 \begin{align}
  \nu^{(\ell)}\ijk &\gets \Round \left\{ - \left( \omega\uk \tau_{ij}^{(\ell)} + \phi\ijk \right) / 2\pi \right\}, \label{eq:upd_mnu} \\
  \theta^{(\ell)}\ijk &\gets \omega\uk \tau_{ij}^{(\ell)} + \phi\ijk + 2\nu^{(\ell)}\ijk\pi, \ \forall i,j,k, \label{eq:upd_mtheta} \\
  \bar{\bt}^{(\ell+1)} &\gets \bar{\bt}^{(\ell)} - \left( \sum_{k=0}^{\T/2} \bar{\bm{C}}^{(\ell)}\uk \right)\invmat \sum_{k=0}^{\T/2} \bar{\bm{c}}^{\prime(\ell)}\uk, \label{eq:upd_mtau}
 \end{align}
 where $\bar{\bm{c}}'\uk = (c'_j)_{k, 1 \leq j \leq M, j \neq {r}}$ and
 \begin{align}
 c'\jk = \omega\uk\sum_{i=1}^{M}B\ijk \frac{\theta\ijk}{\omega\uk}.
 \end{align}
 Note that the above update rules are in complete agreement with those in the case of two sensors \eqref{eq:upd_n}--\eqref{eq:upd_tau}, and hence the \ac{TD} estimate is also identity.


\subsection{Amplitude Estimation}
\label{subsec:est_a}
Although our purpose is to estimate subsample \acp{TD}, we need to also estimate the unknown parameter $\ba_k$.
One of the solutions is the use of fixed amplitudes $\ba_k = \bm{1}$ for all $k$.
However, we here propose the following method for estimating $\ba_k$, which may result in improved \ac{TD} estimates.
\par
First, we arrange the numerator of $\J\uk\left(\ba\uk, \bt\right)$ as
\begin{align}
  \RTFh \bV\uk \, \RTF &=  \ba\uk\tp \bV\uk'\left(\bt\right) \, \ba\uk, \\
 \bV\uk'\left(\bt\right) &= \bV\uk \odot \bm{P}\uk\left(\bt\right), \label{eq:Vp} \\
  \bm{P}\uk\left(\bt\right) &= \bm{p}\uk\left(\bt\right) \bm{p}\uk\htp\left(\bt\right), \label{eq:P} \\
  \bm{p}\uk\left(\bt\right) &= \vectorize{e^{\iu \omega\uk \tau_{r1}}}{e^{\iu \omega\uk \tau_{rM}}}\tp, \label{eq:p}
\end{align}
where $\bV'\uk\left(\bt\right)$ is an Hermitian matrix and $\bm{P}\uk\left(\bt\right)$ is an $M \times M$ matrix\footnote{Properties of these matrices are discussed in the appendix.}.
At fixed $\bt$, the optimization problem \eqref{eq:opt_maux} becomes
\begin{gather}
 \argmax_{\ba\uk} \ \frac{1}{\T}\sum_{k=0}^{\T/2} \beta\uk \J\uk(\ba\uk) \ \ \mathrm{s.t.} \ \ a_{mk} \geq 0\ \ \forall m, k, \label{eq:obj_a} \\
 \J\uk(\ba\uk) = \frac{\ba\uk\tp \bV'\uk \ba\uk}{\ba\uk\tp \ba\uk} = \frac{\ba\uk\tp \Re[\bV'\uk] \ba\uk}{\ba\uk\tp \ba\uk}, \label{eq:rayleigh}
\end{gather}
where $\Re[\cdot]$ takes the real part of the input argument, and again $\beta_0 = \beta_{\T/2} = 1$ and $\beta_k=2$ for $k\not\in \{0,\T/2\}$.
The objective function \eqref{eq:rayleigh} is known as the {\it Rayleigh quotient}, where $\ba_k$ is the non-zero (and also non-negative in our problem) vector.
Hence, the optimization problem is the maximization of the Rayleigh quotient with a non-negative constraint, which may be of general interest.

\subsubsection{Unconstrained case}
\label{subsubsec:eig}
The solution of this maximization problem without any constraint is given by the eigenvalue decomposition (e.g., \cite{Kenneth2016}).
Since the objective function \eqref{eq:rayleigh} is invariant to the scale of the amplitudes, we optimize $\ba\uk$ under the constraint $\|\ba\uk\|_2^2 = 1$ and compensate for $a_{rk}$ to $1$ at the end of update sequences.
Then, using the method of Lagrange multipliers, we find the stationary point by taking the gradient with respect to $\ba\uk$ and setting it to zero as
\begin{align}
 \frac{\partial \J\uk(\ba\uk)}{\partial \ba\uk} = 
 \bV'\uk \ba\uk - \lambda\uk \ba\uk &= \bm{0}, \label{eq:eig}
\end{align}
where $\lambda\uk$ is the $k$th Lagrange multiplier.
\par
This result is well known as the eigenvalue problem.
Although any eigenvector of $\bV\uk'$ used as $\ba\uk$ satisfies \eqref{eq:eig}, we choose the eigenvector corresponding to the largest eigenvalue (hereafter, we simply denote it as the largest eigenvector) for maximizing the objective function \eqref{eq:rayleigh}.

\subsubsection{Constrained case}
One solution for the maximization of the Rayleigh quotient with non-negative constraint is the projection.
Since the largest eigenvector can take a negative value, we project its elements $\hat{a}_{mk}$ to the positive domain
\begin{align}
 \hat{a}_{mk} \leftarrow \max(\hat{a}_{mk}, \ 0) \ \ \forall m \label{eq:projection}
\end{align}
to satisfy the non-negative constraint.
\par
Although the largest eigenvector is absolutely the solution of \eqref{eq:rayleigh}, there is no guarantee that the projected eigenvector maximizes the objective function. 
We thus propose the alternative method based on the auxiliary function method.
\begin{theorem}
\label{th:a}
Let $\|\ba\uk\|_2^2 = 1$ for all $k$. The following is an auxiliary function for the Rayleigh quotient $\J\uk(\ba\uk)$ \cite{Kenneth2016},
\begin{align}
 Q^a\uk(\ba\uk, \btheta) &= 2\btheta\uk \bV\uk' \ba\uk - \btheta\uk \bV'\uk \btheta\uk, \label{eq:Qa}
\end{align}
 where $\btheta = \vectorize{\btheta_0}{\btheta_{\T/2}}\tp$ are auxiliary variables and $Q^a\uk(\ba\uk, \btheta) = \J\uk(\ba\uk)$ holds when
\begin{align}
 \btheta\uk = \ba\uk.
\end{align}
\end{theorem}
The auxiliary function \eqref{eq:Qa} is the linear form of the vector $\ba\uk$ and is easily maximized even under the non-negative constraint.
Then, the update rules are
\begin{align}
 \ba\uk^{(\ell+1)} &\gets \bV\uk' \ba\uk^{(\ell)}, \label{eq:upd_a}\\
 a_{mk}^{(\ell+1)} &\gets \max\left(a_{mk}^{(\ell+1)}, \ 0\right) \ \ \forall m, \label{eq:max_a}\\
 \ba\uk^{(\ell+1)} &\gets \ba\uk^{(\ell+1)} / \|\ba\uk^{(\ell+1)}\|_2^2. \label{eq:norm_a}
\end{align}
These update sequences are guaranteed to converge to a local maximum, whereas the projected largest eigenvector \eqref{eq:projection} does not attain it unless all elements are positive without projection.
\par
Interestingly, there is a well-known algorithm, namely, the {\it power method} (power iteration), for estimating the largest eigenvector.
The power method iteratively updates the eigenvector estimate in \eqref{eq:upd_a} and \eqref{eq:norm_a}.
The above algorithm shows that we can obtain the local maximum in the same scheme even under the non-negative constraint by applying the projection to the positive domain \eqref{eq:max_a}.

\subsubsection{Shared amplitude case}
\label{subsubsec:ave}
We can consider the observation model with a shared amplitude vector for all frequencies, that is, $\ba\uk$ is identical for all $k$.
The amplitude estimation with this alternative model can be easily realized by using averaged $\bV\uk'$:
\begin{align}
 \bV'_{{\rm mean}} = \frac{1}{\T/2 + 1} \sum_{k=0}^{\T/2} \bV\uk',
\end{align}
in update sequences \eqref{eq:upd_a}--\eqref{eq:norm_a}.
We expect that this alternative model is robust against the estimation error in amplitudes and measurement environments.

\subsection{Update of Noise Variances}
To compute the observations and \ac{RTF} weighted by the standard deviation $\sigma_m$ defined in \eqref{eq:xs} and \eqref{eq:gs}, respectively,
we finally update the variance $\bm{\Sigma}$ of the complex multivariate Gaussian distribution $\mathcal{N}_c\left(\bm{\mu}, \bm{\Sigma}\right)$ as follows:
\begin{align}
\sigma_m^2 \gets \frac{1}{\N\T} \sum_{k=0}^{\T/2} \sum_{n=0}^{\N-1} \beta_k|x\kn[m] - \hat{\bigs}\kn g_{mk}\left(a_{mk}, \tau_{rm}\right)|^2, \label{eq:upd_sigma}
\end{align}
where $g_{mk}\left(a_{mk}, \tau_{rm}\right) = a_{mk}e^{-\jwkt_{rm}}$ and $\hat{\bigs}\kn$ is computed by \eqref{eq:s}.

\subsection{Algorithm of AuxTDE}
Finally, we summarize the update sequences of the \ac{AuxTDE} in \algref{alg:auxtde}, where three types of iteration exist, for the update of $\bt$, $\ba\uk$, and their alternate updates indexed by $\ell_{\tau}$, $\ell_a$ and $\ell_e$, respectively.
The maximum iterations $L_{\tau}$, $L_{a}$, and $L_{e}$ can be $1$.
The initialization of $\bt$ can be performed by the pairwise method.
For instance, the discrete maximum (GCC method \cite{knapp76}) and the result of parabolic interpolation \cite{giovanni93} can be used.
Basically, better initial estimates lead to faster convergence, and parabolic interpolation is thus better in practice.

\begin{figure}[t]
 \begin{algorithm}[H]
  \caption{AuxTDE}
  \label{alg:auxtde}
  \begin{algorithmic}
   \Require $\bx\kn$
   \Ensure $\bt$
   \Initialize{
   Initialize $\bt$\\
   Initialize $\ba\uk = 1$ for all $k$ \\
   Initialize $\sigma_m^2 = 1$ for all $m$ \\
   Compute $\phi\ijk$ and $A\ijk$ for all $i, j, k$ from $\bV\uk$
   }
   \For{$\ell_e = 0, 1, \dots, L_e-1$}
   \State $A'\ijk \gets A\ijk / \sigma_i\sigma_j$ \algorithmiccomment{corresponds to \eqref{eq:xs}}
   \For{$\ell_{\tau} = 0, 1, \dots, L_{\tau}-1$}
   \State $\bt \gets {\tt update\_td}(\ba\uk, \bt, A'\ijk, \phi\ijk)$
   \EndFor
   \For{$\ell_a = 0, 1, \dots, L_a-1$}
   \State $\ba\uk \gets {\tt update\_amp}(\ba\uk, \bt, A'\ijk, \phi\ijk)$ for all $k$
   \EndFor
   \State Update $\sigma_m^2$ by \eqref{eq:upd_sigma}
   \EndFor
   \Lfunction{${\tt update\_td}$}{$\ba\uk$, $\bt$, $A\ijk$, $\phi\ijk$}{
   Update $\btheta$ by \eqref{eq:upd_mnu} and \eqref{eq:upd_mtheta} \\
   Compute $\bm{C}\uk$ and $\bm{c}\uk$ by \eqref{eq:C}--\eqref{eq:c} \\
   Update $\bt$ by \eqref{eq:upd_mtau} \\
   return $\bt$
   }
   \Lfunction{${\tt update\_amp}$}{$\ba\uk$, $\bt$, $A\ijk$, $\phi\ijk$}{
   Compute $V\uk'$ by \eqref{eq:Vp}--\eqref{eq:p} \\
   Update $\ba\uk$ by \eqref{eq:upd_a}--\eqref{eq:norm_a} \\
   \State return $\ba\uk$
   }
  \end{algorithmic}
 \end{algorithm}
\end{figure}


\section{Experimental Analysis of the AuxTDE algorithm}
\label{sec:exp1}

\subsection{Empirical Convergence to the Local Maximum}
\label{subsec:conv1}
\begin{figure}[t]
 \begin{minipage}{0.53\hsize}
  \centering
  \includegraphics[width=1\columnwidth]{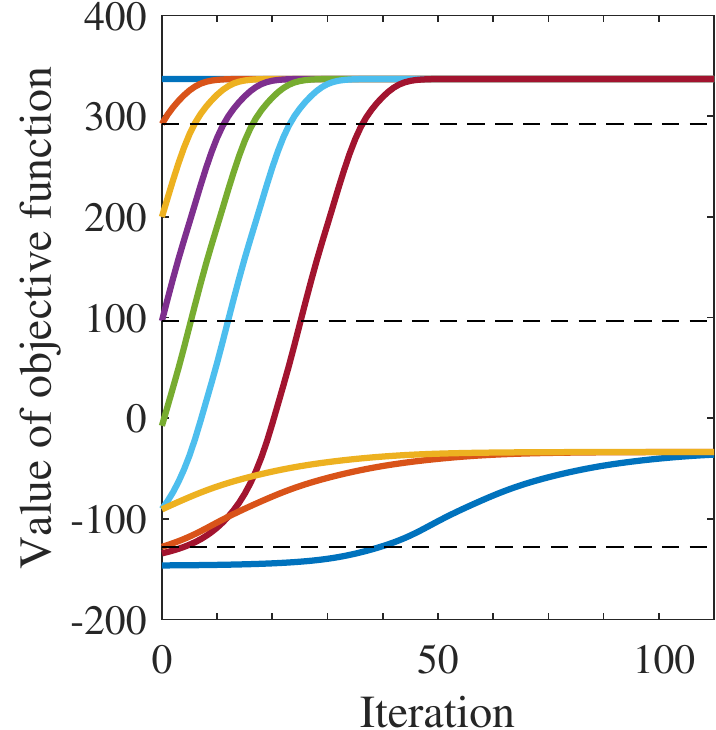}
  \vspace{4pt}
 \end{minipage}
 \begin{minipage}{0.45\hsize}
  \centering
  \includegraphics[width=1\columnwidth]{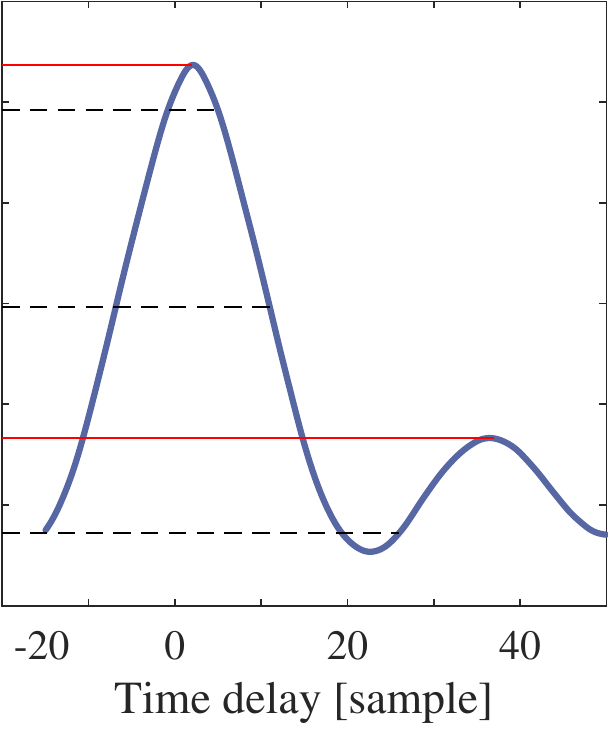}
  \vspace{0pt}
  \end{minipage}
  \vspace{-18pt}
 \caption{Left: the value of objective function via the proposed method over initial values, where $M = 2$. Different curves represent different values of the initial TD estimate, taken every three samples from $2$ to $29$. Right: objective function.}
 \label{fig:init}
\end{figure}

First, we confirm the convergence of the \ac{AuxTDE} by depicting the objective function, where we simulated $M=2$ observations whose sub-sample \ac{TD} was set to $2.0996$ samples.
We used English female speech as a target signal sampled at \kHz{16} and added white Gaussian noise to each microphone with \ac{SNR} of \dB{10}.
\figref{fig:init} shows the convergence of the objective function \eqref{eq:obj_maux} with the \ac{AuxTDE} for different initial values.
We set the initial value of the proposed method to every three samples from $2$ to $29$.
In accordance with this figure, we can confirm that the proposed method will converge to the global maximum if an appropriate initial value is given.
Moreover, this figure shows the guaranteed monotonic increase in the objective function for any initial value.
The initial value must be picked from the unimodal period, including the global maximum, to reach it, where the range is between $-20$ to $22$ samples in this figure.
Basically, the conventional \ac{GCC} method is a good way to obtain such an initial estimate.
Even when the initial estimate is outside the appropriate range, the convergence to the local maximum is always guaranteed owing to the characteristic of the auxiliary function method.
Additionally, the better the initial estimate is (e.g., in the case of using the parabolic interpolation), the faster the convergence is.

\subsection{Empirical Convergence and Consistent TD Estimates}
\label{subsec:conv2}
\begin{figure}[t]
 \centering
 \includegraphics[width=0.8\columnwidth]{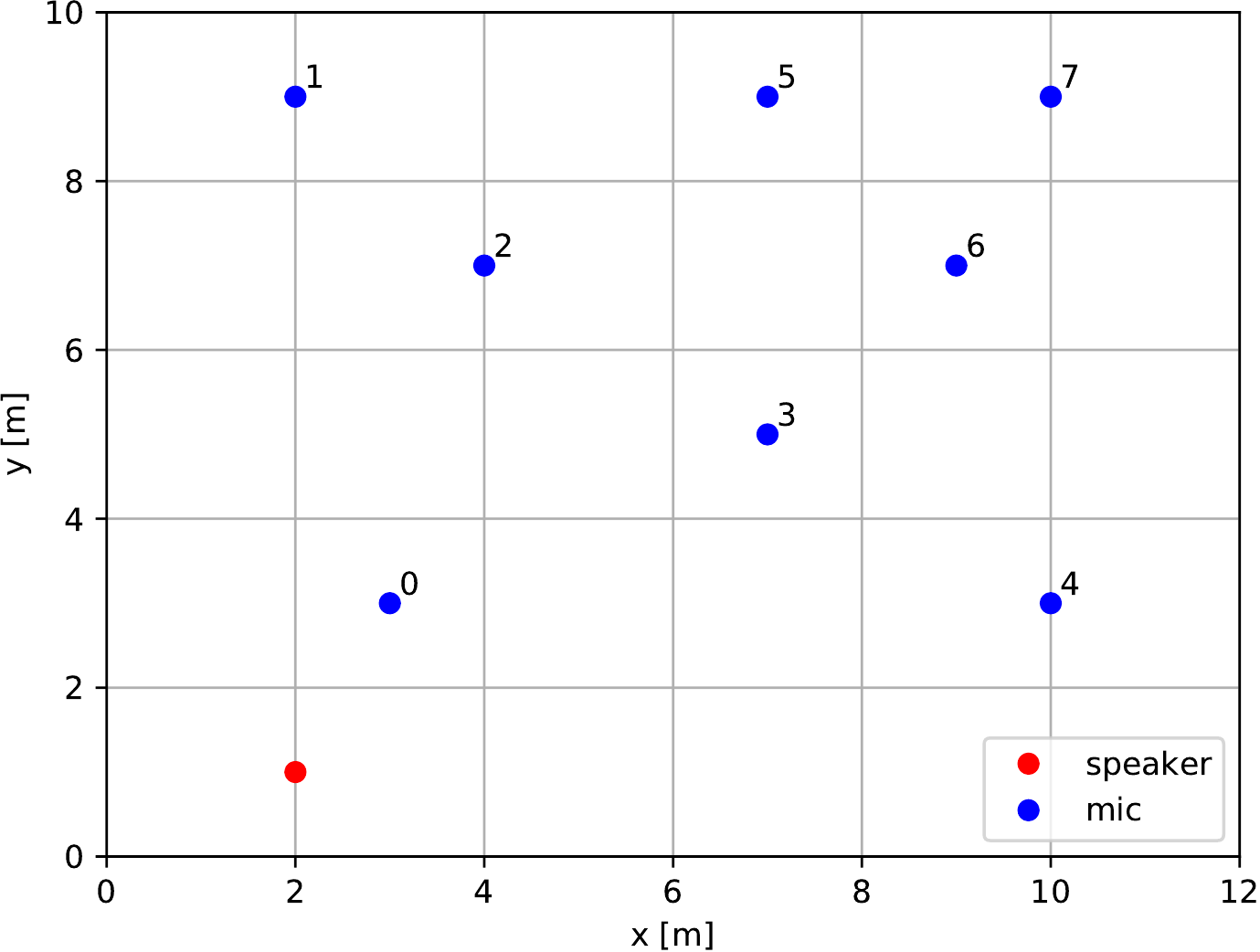}
 \caption{Example of the microphones and target source alignment. The numbers near the microphones denote their indices.}
 \label{fig:room}
\end{figure}

\begin{figure}[t]
 \centering
 \includegraphics[width=1\columnwidth]{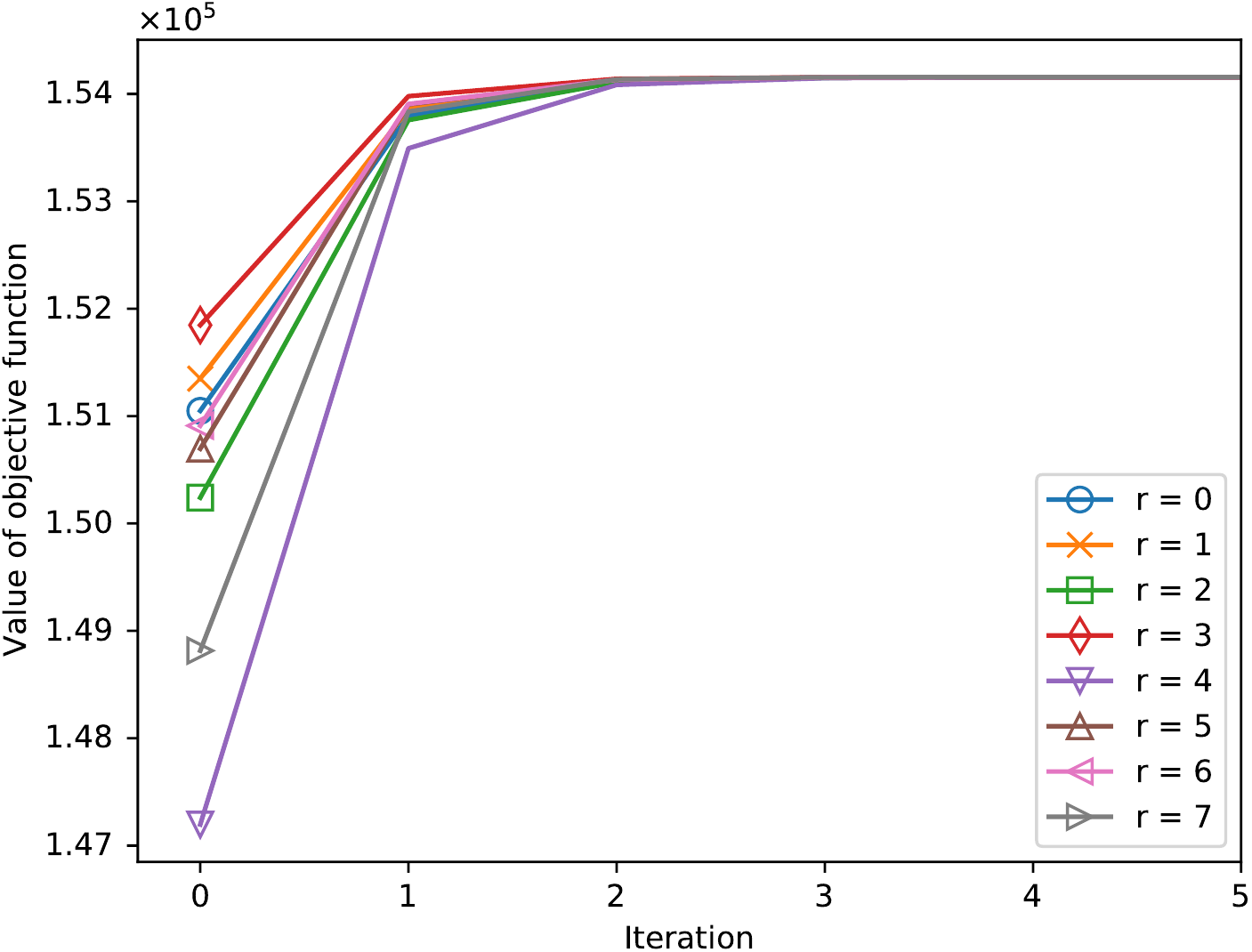}
 \caption{Objective function and convergence via the iterative updates of the AuxTDE. The $0$th iteration corresponds to the objective function using the initial value $\bt^{(0)}$ estimated by the pairwise GCC method with parabolic interpolation. Different curves correspond to the different reference microphones specified by $r$.}
 \label{fig:consistent}
\end{figure}

As we mentioned in \secref{subsec:consistency}, the \ac{AuxTDE} attains consistent \ac{TD} estimates owing to the observation model.
Although we must set the reference sensor for computing \acp{TD} since these are relative values, the \ac{AuxTDE} can obtain the same \ac{TD} estimates regardless of the reference sensor.
This property can be confirmed by verifying the objective function.
Here, we used \textit{pyroomacoustics} \cite{scheibler2018pyroomacoustics} to simulate a reverberant environment and generated eight microphone signals with a target source, as shown in \figref{fig:room}.
In this experiment, we focused on the \ac{TDE} algorithm of the \ac{AuxTDE} and fixed the amplitude $\ba_k$ and the variance $\sigma^2_m$ to $1$ for all elements.
We set initial values of $\bt$ as the estimates by the pairwise method using the parabolic interpolation \cite{giovanni93} of the \ac{GCC} method \cite{knapp76} whose weights were one for all frequencies.
For the pairwise method, there were eight choices of the reference microphone $r$; thus, we computed the same number of the \ac{TD} estimate vector $\hat{\bt}_{r}$.
Then, we computed the objective function of the \ac{AuxTDE} \eqref{eq:obj_maux} with every $\hat{\bt}_{r}$ and showed them in \figref{fig:consistent} at the $0$th iteration.
\par
The objective function values at the $0$th iteration in \figref{fig:consistent} show the dependence of the pairwise \ac{GCC} method on the reference microphone $r$.
This result implies that the performance of \ac{TD} estimation also varied.
Interestingly, the best performance in terms of the objective function was achieved with $r=3$, whereas the worst one was achieved with $r=4$ (see \figref{fig:room}).
It is difficult to predict the best reference microphone in advance, even when the actual layout of the source and microphones was known.
Moreover, since we cannot obtain true \ac{TD} values in practice, the evaluation criterion for the pairwise method is also unclear.
\par
In contrast, the \ac{AuxTDE} with these initial estimates monotonically increased the objective function and converged to the same local maximum regardless of the reference microphone, as shown in \figref{fig:consistent}.
This result indicates that consistent \acp{TD} were obtained owing to the joint optimization \eqref{eq:opt_maux} and the transfer system model \eqref{eq:RTF}.
Additionally, it can be said that these results are the best in terms of \ac{ML}.
Similar to the result in \subsecref{subsec:conv1}, the better initial estimate led to faster convergence.

\section{Experiments of Time Delay Estimation}
\label{sec:exp2}


\subsection{Experimental Condition}
\label{subsec:exp_cond}
In this section, we evaluated the performance of \ac{TD} estimation.
We used the pyroomacoustics \cite{scheibler2018pyroomacoustics} to simulate reverberant room environments.
We synthesized $M$ observed signals sampled at \kHz{16} with simulated \aclp{RIR} with a reverberation time of approximately \ms{200}.
The target signal of \s{5} was randomly generated following normal Gaussian distribution and was contaminated by additive Gaussian noise, where \ac{SNR} was set to \dB{20}.
The target source and microphones are randomly located in a room of \m{3} $\times$ \m{4} $\times$ \m{3} size, as in the example shown in \figref{fig:room}.
We tested three types of microphone alignment: widely placed $M=4$ and $M=8$ microphones assuming \acp{DMA}, where they are located at least \m{0.2} away from each other, and closely placed $M=8$ microphones assuming an ordinary microphone array.
Note that we assumed that all microphones are synchronized, and no \acl{SFM} problem \cite{Alexander11} occurred.
We performed \ac{STFT} with a rectangle window for the observed signals, where the window length is $4096$ samples, and each frame is half-overlapped.
\par
We evaluated three types of \ac{AuxTDE} algorithm for amplitude estimation: the original \ac{AuxTDE}, denoted as AuxTDE\_freqAmp, estimates both \acp{TD} and frequency-dependent amplitudes simultaneously as shown in \algref{alg:auxtde}, AuxTDE\_shrdAmp estimates shared (frequency-independent) amplitude by the algorithm described in \subsecref{subsubsec:ave}, and AuxTDE\_unitAmp estimates only \acp{TD} with fixed amplitudes ($\ba_k = \bm{1} \ \forall k$).
The number of iterations of each method is listed in \tblref{tbl:iters}, and all the methods update the variances.
\par
For comparison, we evaluated three types of pairwise methods: the \ac{GCC} method \cite{knapp76} (PW-GCC), the \ac{GCC} method with parabolic interpolation \cite{giovanni93} (PW-Parafit), and the \ac{GCC} method with the \ac{AuxTDE} for two channels (PW-AuxTDE).
Although the \ac{AuxTDE} is applicable with three or more microphones, we used it here as the pairwise method merely for comparison.
We used every microphone as the reference one for all the methods and obtained $M$ \ac{TD} estimate vectors $\hat{\bt}_r$.
\par
For the evaluation criterion, we used \acp{RMSE} between the estimated \acp{TD} $\hat{\bt}_{r,p}$ and computed ones $\bt\gt_{r,p}$ defined as\footnote{The \ac{TD} estimate at the reference microphone $\tau_{rr}$ is always zero, and the number of \ac{TD} estimates is thus $M-1$.}
\begin{align}
 {\rm RMSE} = \sqrt{\frac{1}{PM(M-1)}\sum_{p=1}^{P}\sum_{r=1}^{M}\|\hat{\bt}_{r,p} - \bt\gt_{r,p}\|^2_2},
\end{align}
where $P$ denotes the number of simulations, and the subscript $p$ denotes the simulation index.
$P$ was $1200$ in this experiment.
Since the true \acp{TD} were unknown, we computed the \acp{TD} from the distance between the target source and each microphone and used it as the ground truth instead, where the speed of sound was \SI{343}{m/s} (pyroomacoustics default).
Note that we eliminated several gross error cases from the evaluation. 
The occurrence of gross errors depended on the conditions (e.g., microphone positions and initial estimates) and was approximately $\pct{0.1}$ in this experiment.
\par
In addition to the evaluation of \ac{RMSE}, we evaluate how inconsistent the \ac{TD} estimates are.
On the basis of \defref{def:consistency} for the consistent \acp{TD}, we define the \ac{MID} as follows\footnote{When $r = r'$, $\left|\tau_{r'm} - \left(\tau_{r'r} + \tau_{rm} \right)\right|$ is always zero for all $m$, and we thus eliminated this case from the parameter.}:
\begin{equation}
 \mathrm{MID} = \left(\frac{1}{M-1}\right)^{\!\!2} \sum_{r'=1}^{M}\sum_{m=1}^{M} \bigl[\left|\tau_{r'm} - \left(\tau_{r'r} + \tau_{rm} \right)\right|\bigr],
\end{equation}
where $r=1$ in this experiment.
Clearly, \ac{MID} is $0$ if the \ac{TD} estimates are completely consistent; otherwise, it takes a high value.
To evaluate the \ac{MID}, we thus need to perform \ac{TD} estimation $M$ times in total by setting each sensor as the reference one.

\begin{table}[t]
 \centering
 \caption{Number of iterations of each AuxTDE variant.}
 \label{tbl:iters}
  \begin{tabular}{@{}lccc@{}} \toprule
   Method & $\ell_{\tau}$ & $\ell_a$ & $\ell_e$ \\ \midrule
   AuxTDE\_freqAmp & 10 & 10 & 3 \\
   AuxTDE\_unitAmp & 10 & - & 3 \\
   AuxTDE\_shrdAmp & 10 & 10 & 3\\ \bottomrule
  \end{tabular}
\end{table}

 \begin{table}[t]
  \centering
  \caption{RMSEs between the estimated and ground truth \acp{TD} for each method and environment.}
  \label{tbl:results}
  {\tabcolsep = 1.6mm
  \begin{tabular}{@{}lccc@{}}
   \toprule
   Method & \multicolumn{3}{c}{\# of microphones $M$} \\ \cmidrule(l){2-4}
   & 4 (DMA) & 8 (DMA) & 8 (Array) \\ \midrule
   PW-GCC & $28.8 \times 10^{-2}$ & $28.9 \times 10^{-2}$ & $28.8 \times 10^{-2}$ \\
   PW-Parafit & $8.45 \times 10^{-2}$ & $8.44 \times 10^{-2}$ & $8.56 \times 10^{-2}$ \\
   PW-AuxTDE & $1.40 \times 10^{-2}$ & $1.38 \times 10^{-2}$ & $2.21 \times 10^{-2}$ \\  \midrule
   AuxTDE\_unitAmp & $1.07 \times 10^{-2}$ & $0.826 \times 10^{-2}$ & $\mathbf{1.37 \times 10^{-2}}$ \\
   AuxTDE\_freqAmp & $1.14 \times 10^{-2}$ & $0.856 \times 10^{-2}$ & $1.48 \times 10^{-2}$ \\
   AuxTDE\_shrdAmp & $\mathbf{1.06 \times 10^{-2}}$ & $\mathbf{0.817 \times 10^{-2}}$ & $1.39 \times 10^{-2}$ \\  \bottomrule
  \end{tabular}
  }
 \end{table}


 \begin{table}[t]
  \centering
  \caption{\acp{MID} for each method and environment.}
  \label{tbl:results_ic}
  {\tabcolsep = 1.6mm
  \begin{tabular}{@{}lccc@{}}
   \toprule
   Method & \multicolumn{3}{c}{\# of microphones $M$} \\ \cmidrule(l){2-4}
   & 4 (DMA) & 8 (DMA) & 8 (Array) \\ \midrule
   PW-GCC & $16.5 \times 10^{-2}$ & $21.7 \times 10^{-2}$ & $21.4 \times 10^{-2}$ \\
   PW-parafit & $6.68 \times 10^{-2}$ & $8.71 \times 10^{-2}$ & $8.83 \times 10^{-2}$ \\
   PW-AuxTDE & $1.09 \times 10^{-2}$ & $1.40 \times 10^{-2}$ & $2.22 \times 10^{-2}$ \\ \midrule
   AuxTDE\_unitAmp & $3.88 \times 10^{-12}$ & $3.88 \times 10^{-12}$ & $3.73 \times 10^{-12}$ \\
   AuxTDE\_freqAmp & $2.85 \times 10^{-12}$ & $0.941 \times 10^{-12}$ & $7.79 \times 10^{-12}$ \\
   AuxTDE\_shrdAmp & $0.162 \times 10^{-12}$ & $0.122 \times 10^{-12}$ & $0.290 \times 10^{-12}$ \\
   \bottomrule
  \end{tabular}
  }
 \end{table}

\begin{figure*}[t]
 \centering
 \subfloat[]{\includegraphics[width=0.47\linewidth]{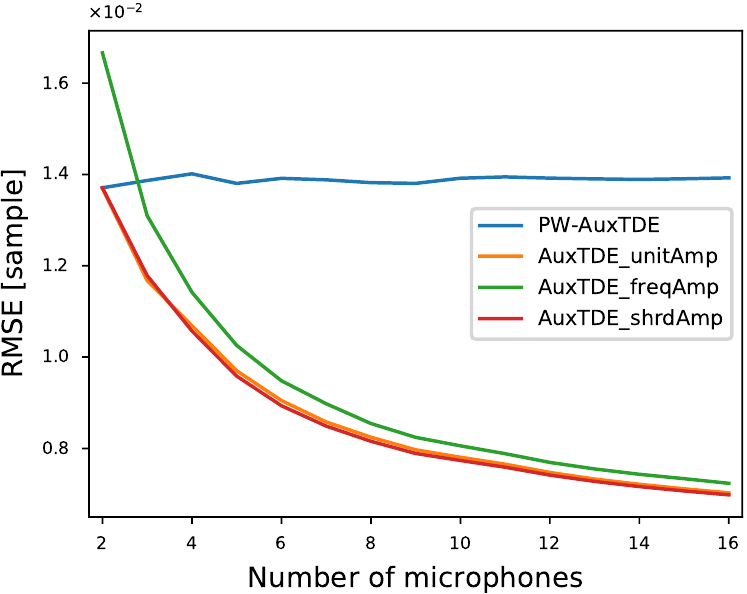}\label{subfig:rmse}}\quad %
 \subfloat[]{\includegraphics[width=0.47\linewidth]{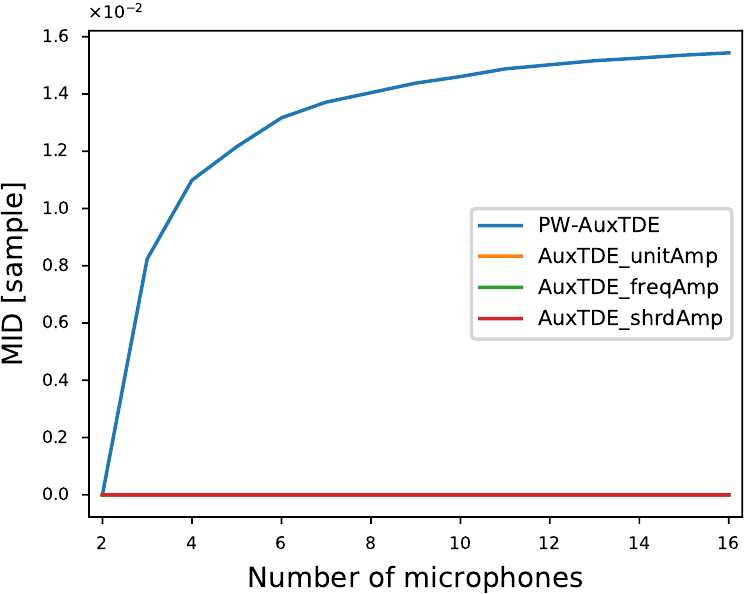}\label{subfig:ic}}%
 \caption[]{(a) Accuracy of \ac{TD} estimation and (b) \acp{MID} in \ac{TDE} as functions of the number of microphones $M$, where $M$ varies in the range of $2$ to $16$. The \acp{MID} of AuxTDE algorithms except for PW-AuxTDE varied from $10^{-12}$ to $10^{-13}$. The proposed methods benefited from using more microphones, while the pairwise method did not.}
 \label{fig:results_mic}
\end{figure*}

\subsection{Results and Discussion}
\label{subsec:results}
\tblref{tbl:results} shows the \acp{RMSE}, the results of \acp{TD} estimation.
The theoretical error in PW-GCC is $0.25$, and values close to this error were obtained.
PW-Parafit significantly improved the estimation accuracy with quite low computational cost, and PW-AuxTDE attained greater improvement with the efficient iterative algorithm.
Their \acp{RMSE} were almost the same for all the microphone alignments.
\par
The \acp{RMSE} of AuxTDE algorithms except for PW-AuxTDE were superior to that of the pairwise methods in every case.
Additionally, the performance was improved by using eight microphones than four microphones.
Since our model includes spatial information contained in the entire observation, AuxTDE could utilize the consistency in \ac{TD} parameters.
Furthermore, the performance with \acp{DMA} was relatively higher than in the case of using an array of closely placed microphones in this experiment.
This result implies that one of the suitable applications of the \ac{AuxTDE} is a \ac{DMA} (and other distributed sensor systems), which consists of widely placed sensors and has broad spatial information.
\par
The simultaneous estimation of the amplitude (AuxTDE\_freqAmp) degraded the performance in most cases compared with AuxTDE\_unitAmp even though the performance of AuxTDE\_freqAmp was superior to that of the pairwise methods.
As one reason, we can consider that the constant amplitude used in AuxTDE\_unitAmp (i.e., $\ba\uk = \bm{1}$ for all $k$) was an excellent a priori for \ac{TD} estimation.
Moreover, it is possible that estimating frequency-dependent amplitudes overfitted the acoustic environment.
Estimating shared amplitudes (AuxTDE\_shrdAmp) may be the better solution for some situations such as when using \ac{DMA}.
For example, when the gain of each sensor differs, the mechanism of AuxTDE\_shrdAmp may be able to reduce the negative effect due to their gain differences.
\par
\tblref{tbl:results_ic} shows the \ac{MID} of the \ac{TD} estimates.
The \acp{MID} of the pairwise methods were considerably high, and the order of \acp{MID} was the same as the \acp{RMSE} (see also \tblref{tbl:results}).
This means that the \ac{TD} estimates were inconsistent.
Therefore, there should be one best microphone that should be used as the reference microphone; however, the method to find it is unclear.
On the other hand, the proposed methods that estimate all \acp{TD} simultaneously showed markedly low \acp{MID} regardless of the microphone alignment.
This means that the \ac{TD} estimates were independent of the reference microphone.
From the above, we can confirm the effectiveness of the proposed \ac{AuxTDE} for \ac{TD} estimation.

\subsection{TD Estimation with a Number of Microphones}
\label{subsec:results_mic}
Finally, we investigated the relationship between the accuracy in \ac{TDE} and the number of microphones to compare the proposed methods and the pairwise methods further.
Experimental conditions are identical to those described in \subsecref{subsec:exp_cond} except for the number of microphones $M$.
It varies from $2$ to $16$, and they are located at least \m{0.2} away from each other, assuming \acp{DMA}.
\par
\figsref{fig:results_mic}\subref{subfig:rmse} and \subref{subfig:ic} show the \ac{RMSE} and \ac{MID} of each method as functions of the number of microphones, respectively.
Note that the results of PW-GCC and PW-Parafit are omitted because their performance changes with respect to $M$ tended to be the same as that of PW-AuxTDE.
The \ac{TD} estimates of each method except for AuxTDE\_freqAmp are theoretically identical when $M=2$.
\par
From \figref{fig:results_mic}\subref{subfig:rmse}, the \ac{RMSE} of PW-AuxTDE was independent of the number of microphones $M$.
Since the pairwise method only uses partial spatial cue between two selected sensors, there are no benefits of increasing the number of sensors.
In contrast, the \ac{MID} increased (i.e., worsened) with increasing $M$, as shown in \figref{fig:results_mic}\subref{subfig:ic}.
The \ac{MID} is theoretically zero when $M=2$ since the interchange of the reference sensor corresponds to the time reversal of the \ac{CC} function.
The choices of reference sensors increased with increasing $M$, and as a result, the \ac{MID} became high.
\par
In contrast to the above results, \ac{AuxTDE} algorithms except for PW-AuxTDE improved the performance of \ac{TDE} with increasing number of microphones.
This result demonstrated the importance and efficacy of using entire spatial information captured by the microphones and consistent constraint for \acp{TD}.
Additionally, the \acp{MID} were considerably low (approximately $10^{-12}$); in other words, the \acp{TD} estimated by the proposed methods were consistent regardless of $M$.
Finally, we concluded that \ac{AuxTDE} algorithms are effective for \ac{TDE}, which attains highly accurate and consistent \ac{TD} estimates.


\section{Conclusions}
\label{sec:conclusion}
In this paper, we proposed AuxTDE, a novel method for \ac{TD} estimation, using the auxiliary function method.
The joint optimization problem for consistent \acp{TD} and amplitudes was considered on the basis of \ac{ML} estimation.
The objective function, \ac{MCC} function, encodes the \ac{CC} function of all sensor pairs, which can thus be considered as the multidimensional extension of the \ac{CC} function.
The \ac{MCC}, which is the nonconvex function, was lower-bounded by the quadratic auxiliary function, and efficient update rules that iteratively maximize the \ac{MCC} were derived.
In experiments, we demonstrated important properties of AuxTDE: monotonic increases in the objective function, convergence to the local maximum, and independence against the reference sensor.
Additionally, we confirmed the efficacy of the AuxTDE through the experiment of \ac{TD} estimation, where the AuxTDE attained highly accurate and consistent \ac{TD} estimates.
The future work includes the online extension of the AuxTDE and the heuristic extension of its algorithm.


\appendices
\section{Proof of Proposition \ref{prop:cos} and Theorems \ref{th:aux} and \ref{th:maux}}
\subsection{Proof of Proposition \ref{prop:cos}}
\proof
Let
\begin{align}
    f(\theta) = \cos \theta+\frac{1}{2}\frac{\sin \theta_0}{\theta_0}\theta^2 - \left(\cos \theta_0 + \frac{1}{2}\theta_0 \sin \theta_0\right).
\end{align}
Then, we have
\begin{align}
    f'(\theta) &= -\sin \theta + \frac{\sin \theta_0}{\theta_0}\theta = -\theta\left(\frac{\sin \theta}{\theta}- \frac{\sin \theta_0}{\theta_0}\right).
\end{align}
We separately consider the following three cases. \smallskip

{\noindent \bf Case 1:} $0<|\theta_0|<\pi$\\
Because $\sin \theta/\theta$ is monotonically decreasing in $0 \leq \theta \leq \pi$,
\begin{align}
    f'(\theta) \left\{
    \begin{array}{cl}
      <0 & (0 \leq \theta < |\theta_0|), \\
      =0 & (\theta=|\theta_0|), \\
      >0 & (|\theta_0| < \theta \leq \pi).
    \end{array}
    \right.
\end{align}
It thus appears that $f(\theta)$ attains its minimum at $|\theta_0|$.
Moreover, $f(\theta_0) = 0$ and thus $f(\theta) \geq 0$ in $0\leq \theta\leq \pi$.
Since $f(\theta)$ is an even function, its minimum value in $-\pi \leq \theta \leq \pi$ is also $0$.
Because $\cos \theta$ is periodic but $-\theta^2$ is not, $f(\theta+2\nu\pi)>f(\theta)$ for any $-\pi \leq \theta \leq \pi$ and integer $\nu\neq 0$.
Therefore, $f(\theta)\geq 0$, with equality if and only if $|\theta|=|\theta_0|$. \smallskip

{\noindent \bf Case 2:} $\theta_0=0$\\
In this case, for $0\leq \theta \leq \pi$, we have
\begin{align}
    f'(\theta) \left\{
    \begin{array}{cl}
      =0 & (\theta=|\theta_0|=0) \\
      >0 & (|\theta_0| < \theta \leq \pi)
    \end{array}
    \right.,
\end{align}
which means $f(\theta)$ takes a minimum value at $f(0)=0$ in $-\pi \leq \theta \leq \pi$. 
Similarly to case $1$, we obtain $f(\theta)\geq 0$, with equality if and only if $\theta=0$. \smallskip

{\noindent \bf Case 3:} $\theta_0=\pi$ or $\theta_0=-\pi$\\
In this case, $f(\theta)=\cos \theta+1$. Therefore $f(\theta)\geq 0$, with equality if and only if $\theta=(2\nu+1)\pi$ for any $\nu\in\mathbb{Z}$.
\QED

\subsection{Proof of Theorems \ref{th:aux} and \ref{th:maux}}
\proof
Because $\cos (\omega_k \tau\uij + \phi\ijk)=\cos(\omega_k \tau\uij + \phi\ijk + 2\nu\ijk\pi )$ with $\nu\ijk\in\mathbb{Z}$,
and because $\beta_k a\ik a\jk A\ijk \geq 0$,
we can apply Proposition~\ref{prop:cos} separately to each term of the sum in \eqref{eq:obj_aux} and \eqref{eq:obj_mcos}.
\QED


\section{Properties of matrices in \subsecref{subsec:est_a}}
 $\bm{P}\uk$ is a positive semidefinite matrix because the following is satisfied for any complex-valued vector $\bm{z}$:
 \begin{align}
  \bm{z}\htp \bm{P}\uk \bm{z} = |\bm{p}\uk\htp \bm{z}|^2 \geq 0.
 \end{align}
 In accordance with the Schur product theorem, $\bV\uk'$, which is the Hadamard product of two positive semidefinite matrices, is also the positive semidefinite matrix.
 Additionally, $\rank(\bm{P}\uk) = 1$ and thus $\rank(\bV\uk') = \rank(\bV\uk)$.
 When we assume that the signal $s\kn$ and noises $\bm{u}\kn$ are uncorrelated, the covariance matrix $\bV\uk$ can be divided into the signal and noise parts as
\begin{align}
 \bV\uk &= \E\left[\bx\kn \bx\htp\kn \right] = |s\kn|^2 \bm{g}\uk\bm{g}\uk\htp + \E[\bm{u}\kn\bm{u}\kn\htp].
\end{align}
Following the subadditivity of the rank of the matrix, $\rank(\bV\uk') \leq \rank(\E[\bm{u}\kn\bm{u}\kn\htp]) + 1$, where $\bm{g}\uk\bm{g}\uk\htp$ is a rank-$1$ matrix.


%



\section*{Acknowledgment}
This work was supported by JSPS KAKENHI grant numbers 20H00613 and 19J20420 and JST CREST grant number JPMJCR19A3, Japan.
The authors would like to thank \mbox{Robin Scheibler} for collaboration in the early stage of this work.

\ifCLASSOPTIONcaptionsoff
  \newpage
\fi



\bibliographystyle{IEEEtran}
\bibliography{IEEEabrv,articles,publications}
\end{document}